\documentclass[preprint]{aastex63}
\usepackage{listings}
\usepackage{amsmath}
\usepackage{amsmath}
\usepackage{mathrsfs}

\newcommand{\reply}[1]{ {\color{black}#1}} 
\newcommand{\replyb}[1]{ {\color{black}#1}}

\newcommand\packageurl{\url{https://github.com/teuben/DataComb}}

\received{2022 December}
\revised{2023 January}
\accepted{2023 February}
\submitjournal{PASP}

\shorttitle{Data Combination}
\shortauthors{Plunkett, Hacar, Moser-Fischer, Petry, Teuben, et al.}
\graphicspath{{./}{figures/}}

\begin{document}

\title{Data Combination: Interferometry and Single-Dish Imaging in Radio Astronomy}

\correspondingauthor{Adele Plunkett}
\email{aplunket@nrao.edu}

\author[0000-0002-9912-5705]{Adele Plunkett}
\affiliation{National Radio Astronomy Observatory, 520 Edgemont Road, Charlottesville, VA 22903, USA}

\author[0000-0001-5397-6961]{Alvaro Hacar} 
\affiliation{Department of Astrophysics, University of Vienna, T{\"u}rkenschanzstrasse 17, 1180, Vienna (Austria)} 
\affiliation{Leiden Observatory, Leiden University, PO Box 9513, 2300-RA Leiden (The Netherlands)} 

\author{Lydia Moser-Fischer}
\affiliation{Argelander-Institut fuer Astronomie, Auf dem Huegel 71, D-53121 Bonn, Germany}

\author[0000-0002-8704-7690]{Dirk Petry}
\affiliation{European Southern Observatory, Karl-Schwarzschild-Strasse 2, 85748 Garching, Germany} 

\author[0000-0003-1774-3436]{Peter Teuben}
\affiliation{Astronomy Department, University of Maryland, 4296 Stadium Dr., College Park, MD 20742, USA}

\author[0000-0001-9504-7386]{Nickolas Pingel}
\affiliation{Department of Astronomy, University of Wisconsin--Madison, 475 N. Charter Street, Madison, WI 53706, USA}
\affiliation{Research School of Astronomy and Astrophysics, The Australian National University, Canberra, ACT 2611, Australia} 

\author[0000-0002-1568-579X]{Devaky Kunneriath}
\affiliation{National Radio Astronomy Observatory, 520 Edgemont Road, Charlottesville, VA 22903, USA} 

\author{Toshinobu Takagi}
\affiliation{Japan Space Forum, 3-2-1 Kandasurugadai, Chiyoda-ku, Tokyo 101-0062, Japan} 

\author{Yusuke Miyamoto}
\affiliation{Department of Electrical and Electronic Engineering, Fukui University of Technology, 3-6-1, Gakuen, Fukui, 910-8505, Japan}\affiliation{National Astronomical Observatory of Japan, 2-21-1 Osawa, Mitaka, Tokyo 181-8588, Japan}

\author[0000-0001-9793-5416]{Emily Moravec}
\affiliation{Green Bank Observatory, P.O. Box 2, Green Bank, WV 24944, USA}

\author[0000-0003-0412-8522]{S\"umeyye Suri}
\affiliation{Department of Astrophysics, University of Vienna, T{\"u}rkenschanzstrasse 17, 1180, Vienna (Austria)}

\author[0000-0001-9662-9089]{Kelley M.~Hess}
\affiliation{Instituto de Astrof\'{i}sica de Andaluc\'{i}a (CSIC), Glorieta de la
Astronom\'{i}a s/n, 18008 Granada, Spain}
\affiliation{ASTRON, the Netherlands Institute for Radio Astronomy, Postbus 2, 7990
AA, Dwingeloo, The Netherlands
} 

\author[0000-0003-2523-4631]{Melissa Hoffman}
\affiliation{Space Telescope Science Institute, 3700 San Martin Drive, Baltimore, MD, 21218, USA}

\author[0000-0002-8472-836X]{Brian Mason}
\affiliation{National Radio Astronomy Observatory, 520 Edgemont Road, Charlottesville, VA 22903, USA}



\begin{abstract}
Modern interferometers routinely provide radio-astronomical images down to sub-arcsecond resolution. 
\reply{However, interferometers filter out spatial scales larger than those sampled by the shortest baselines, which affects the measurement of both spatial and spectral features. Complementary single dish data is vital for recovering the
true flux distribution of spatially resolved astronomical sources with such extended emission.}  In this work, we provide an overview of the prominent available methods to combine single dish and interferometric observations. We test each of these methods in the framework of the CASA data analysis software package on both synthetic continuum and observed spectral datasets. We develop a set of new assessment tools which are generally applicable to all radio-astronomical cases of data combination. Applying these new assessment diagnostics we evaluate the methods' performance and demonstrate the significant improvement of the combined results in comparison to purely interferometric reductions. We provide combination and assessment scripts as add-on material. Our results highlight the advantage to use data combination to ensure high-quality science images of spatially resolved objects. 

\end{abstract}



\section{Introduction}
\label{sec:intro}

\begin{figure}[hbt]
\centering
\includegraphics[width=0.6\textwidth]{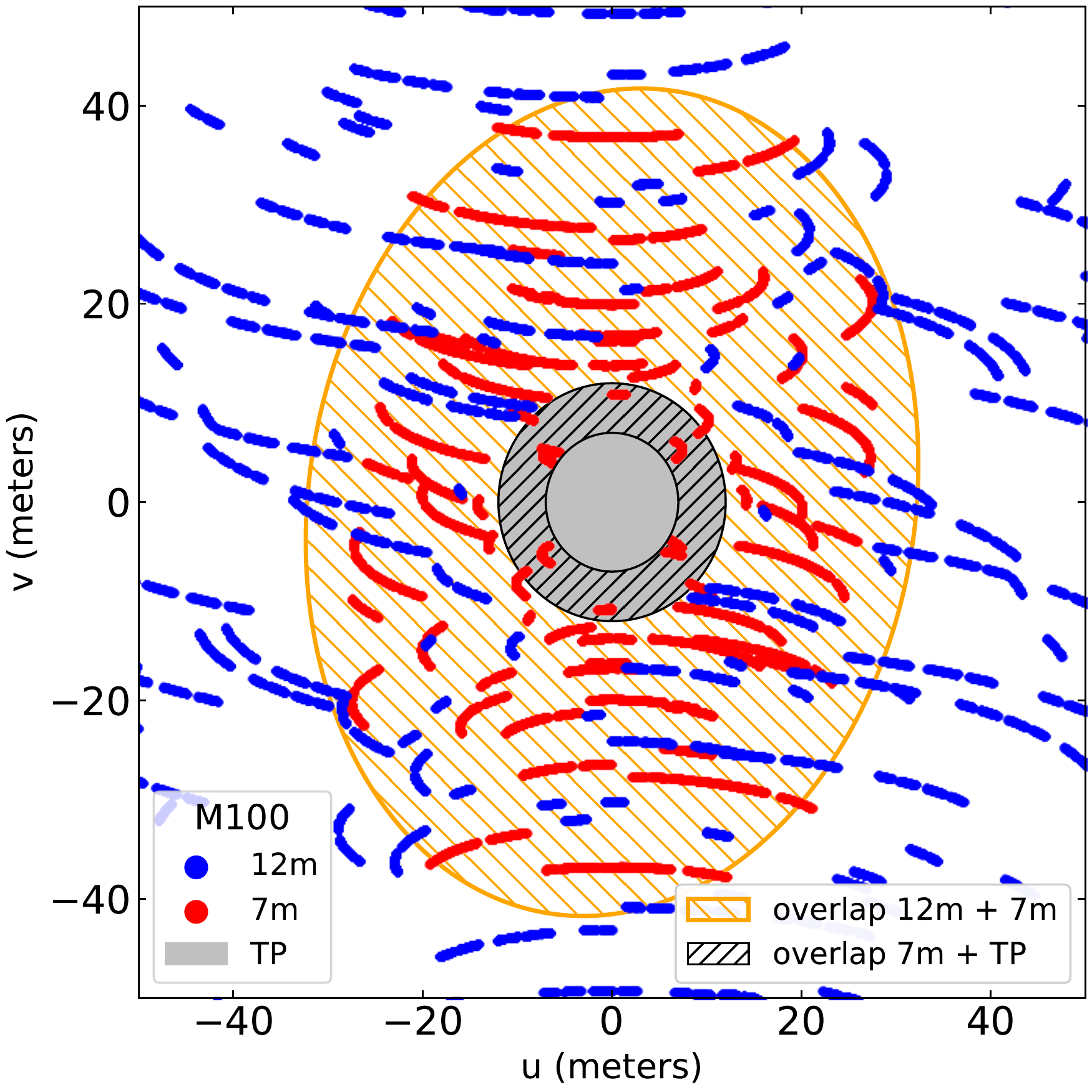}
\caption{\replyb{Central \emph{(u,v)} coverage of the M100 dataset presented in this work showing all baselines of $\leq$~50m observed by the 12m (blue), 7m (red), and Total Power (gray) arrays. Different ellipses highlight the overlapping uv-coverage between the 12m and 7m (hatched orange area) and between the 7m and TP (hatched black area) arrays, respectively. Significant overlap between different arrays is preferred for data combination.}}
\label{fig:uvplot}
\end{figure}

A well-known challenge of interferometric imaging in the radio and sub-millimeter regimes is that it relies on aperture synthesis in which the Fourier or \emph{(u,v)} plane is irregularly or incompletely sampled. While interferometry is a powerful technique for resolving structures on smaller scales than can be attained with a single dish telescope, on the other hand, a single dish telescope is necessary for recovering extended structure. This is commonly known as the ``short-spacing'' problem that we show in Figure \ref{fig:uvplot}, visualized as a ``hole'' in the inner region of the \emph{(u,v)} coverage \reply{\citep[e.g.][]{Wilner1994,Kurono2009}; visualizations in the literature are given by 
\citet{mason2020} (see their \S 2.2 and Figures 2-4) and \citet{PHANGS_datareduction} (Figures 24-25), among others.
} 
The lack of short spacings in the interferometer results in the inability to detect the flux of large spatial scales and thus leads to an incomplete representation of the true sky brightness distribution in the image. \reply{Ideally, observations would sample the complete \emph{(u,v)} space corresponding to the spatial scales under study, and recover consistent amplitudes at overlapping \emph{(u,v)} distances if multiple interferometry configurations and single dish observations are required \citep[see e.g., Figure 3 from][]{koda2019}.} As described in \citet{rau2011}, the impact of the zero-spacing problem is not restricted to continuum images. \reply{Spectra can also be strongly affected, as shown in molecular line observations \citep[e.g.][see also Sect.\ref{sec:lineprofile}]{Pety2013,hacar2018}.}

The solution is to complement the interferometric observations by a separate observation with an instrument sensitive to the large scales, typically a single-dish telescope, using data combination. For example, in the case of the Atacama Large Millimeter/sub-millimeter Array (ALMA), this is done by combining data from the main ALMA-12m array and/or the ALMA-7m Compact Array (ACA) with data from the dedicated ALMA single-dish (SD, or Total Power denoted as TP) telescopes. A broad literature highlights the critical role of data combination in both Galactic and extragalactic studies including a large variety of interferometer plus SD combinations not only within ALMA \citep[for a very thorough example, see][among others]{PHANGS_datareduction} but also using CARMA and NRO 45m/IRAM-30m/FCRAO \citep{koda2011,kong2018,Plunkett2013,Plunkett2015}, SMA and APEX \citep{kauffmann2017}, (E)JVLA and GBT \citep{Pineda2011,Williams2018,THORsurvey}, PdBI/NOEMA and IRAM-30m \citep{Pety2013,Beuther2022}, ALMA-12m and MOPRA/IRAM-30m \citep{Peretto2013,hacar2018}, or ASKAP and Parkes \citep{Pingel2022} telescopes, to cite some examples.

Interferometric image synthesis and data combination is not a new topic. \citet{stanimirovic2002} provides a comprehensive introduction to the problem and the history of solution attempts. An update is given by \citet{mason2020} concentrating on the ``feathering" technique of data combination.
Most generally, methods exist for combination at different steps of the imaging process, either in the spatial frequency (Fourier) domain or in the image domain.  However, a systematic \reply{and quantitative} comparison of the results of these methods is still missing. 

In this work we summarize current data combination techniques and provide new tools to evaluate their performance. After setting the foundation by briefly describing important aspects of imaging purely interferometric data in \S \ref{sec:imaging}, we present the prominent existing methods for combining interferometric and single-dish observations in \S \ref{sec:methods}. 
In \S\ref{sec:testdata}, we specify a simulated and an observed test dataset for our evaluation of methods.
In \S \ref{sec:assess}, we describe in detail our new assessment metrics for interferometric and combined images which are implemented in our new performance evaluation tools. We then run the described combination methods on the test datasets and evaluate the results.
Finally, in section \S \ref{sec:discussion} we derive general recommendations for the use of the combination methods.
The script suite for running the combination methods and our evaluation tools, and the data to reproduce the figures in this paper are available from \packageurl.

\section{Imaging Interferometric Data with CLEAN}  \label{sec:imaging}

\reply{There is a Fourier transform relationship between the signals from the antennas making up an interferometer and the sky brightness distribution. The correlated signals, known as visibilities, are visualised in the $(u,v)$ plane, and an inverse Fourier transform is used to provide an image of the spatial plane \citep[see][]{mason2020}.} 
\reply{The point spread function (PSF) of
the interferometer is the Fourier transform of the visibilities, and as these do not fill the $(u,v)$ plane, the PSF can be highly irregular. \replyb{An initial step of imaging produces the so-called `dirty' image, which is the convolution of
this PSF with the true sky brightness distribution. PSF sidelobes manifest as image artifacts.} Thus, subsequent deconvolution is necessary to optimise image quality. Presently, the most
commonly used deconvolution algorithms are} based on CLEAN \citep{hogbom1974, clark1980}.

CLEAN deconvolves the data from the effects of the PSF of the interferomenter by iteratively modeling the emission as a set of point sources (delta functions). \reply{The final image and residuals are convolved with an idealised restoring beam based on the
FWHM of the PSF.} CLEAN achieves a good image quality for astronomical objects comparable in size to the angular resolution. For objects significantly larger, typically multi-scale (MS) CLEAN \citep{cornwell2008} is used, which employs inverted truncated paraboloids instead of delta functions to model the emission.  \reply{CLEAN algorithms also incorporate options for  weighting visibilities and for interpolating into the missing spacings in the $(u,v)$ plane, which provide a trade-off between resolution and sensitivity, but can themselves introduce artifacts if the sky brightness has structure on the unsampled scales such that the interpolation is inaccurate.}  Further extentions of CLEAN \citep[MT-MFS CLEAN,][]{rau2011} account for the frequency dependence of the \emph{(u,v)} coverage. 

For actual data analysis work in this study, we use exclusively the CASA (Common Astronomy Software Applications) package \citep{casa2022} version 6.1. CASA is the primary data analysis package for ALMA and the Karl G. Jansky Very Large Array (VLA).
Other available radio-astronomical data analysis packages include AIPS \citep{vanmoorsel1996}, GILDAS \citep{GILDASoriginal,GILDAS}, MIRIAD \citep{sault1995} and more recently developed programs such as {\tt wsclean} \citep{offringa2014}, {\tt resolve} \citep{arras2020}, and
{\tt purify} \citep{carrillo2014}.

In CASA, CLEAN and its extensions are all accessible via the \texttt{tclean} task. 
Additionally, all the data combination methods presented in \S \ref{sec:methods} can be implemented by calling CASA tasks. Some are already available as a single CASA task (such as {\tt sdintimaging} and {\tt feather}) or add-on packages for CASA ({\tt tp2vis}).  
Each method in \S \ref{sec:methods} uses \texttt{tclean} (or the underlying libraries of \texttt{tclean}) at some step in the imaging and data combination process. 

The user can control the behaviour of \texttt{tclean} via a large set of input parameters which are explained in the CASA documentation\footnote{\url{https://casadocs.readthedocs.io/}}.
Some of the most critical parameters are those related to (a) the number of deconvolution iterations, (b) the CLEAN threshold (i.e. the minimum flux justifying the placement of a CLEAN component), (c) the spatial masking (where the CLEAN components can be placed), and (d) the handling of the relative weighting of the interferometric visibilities.

To facilitate the comparison among the combination methods, we fix all \texttt{tclean} parameters to common values.
We set the number of iterations to 100,000.  
We use Briggs weighting \citep{briggs1995} with robust value of 0.5 (i.e. balance between natural and uniform weighting) for all runs of {\tt tclean}
as well as {\tt gridder = `mosaic'}; \reply{in this study we deal exclusively with mosaic maps, but {\tt gridder = `mosaic'} applies for the case of combining data from multiple arrays of different antenna sizes, even if they are single pointing observations}.

We explored several masking options, as careful masking is important to mitigate constructive interference of the sidelobes of the synthesized beam \citep{condon1998}. The mask is generated by first utilizing the SD image to mask (include) all regions above 0.3 times the max value in the SD image, and in the next step include any additional regions selected by the ``auto-multithresh'' algorithm in {\tt tclean} \citep{kepley2020}, using the standard ``auto-multithresh'' parameters. This masking technique is determined to capture the regions of larger-scale emission (as in the SD image), as well as any regions of more compact emission that might appear primarily in the interferometry data.
Finally, the spectral definition mode can be chosen to be either `mfs' (multi-frequency synthesis,  for continuum imaging with only one output image channel) or `cube' (for spectral line imaging with one or more channels), depending on the dataset and desired image.  In this manuscript, we present continuum and spectral data, as an image and cube, respectively.


\section{Combination Methods} \label{sec:methods}
In this section, we describe the theory and execution of the currently available data combination methods which we have selected for evaluation in this study (see also Table \ref{tab:methods}). \reply{The methods incorporate one or more of the following steps in order to combine the interferometry dataset(s) with SD observations: combine the SD data with a clean interferometer image; combine the SD data with the dirty interferometer image before joint deconvolution; convert the single-dish data to pseudo-visibilities, combine in the ($u,v$) plane and then FT the combined data and CLEAN.} The methods we present for evaluation are: Feather (\S \ref{sec:feather}), SDINT (\S \ref{sec:sdintimaging}), a model-assisted CLEAN plus Feather (hereafter MACF) (\S \ref{sec:startmodel}), TP2VIS (\S \ref{sec:tp2vis}), and \citet{faridani2018}'s Short Spacing Correction (hereafter FSSC) (\S \ref{sec:faridani}). For each of these methods we include a list of the control parameters. 

\reply{We note that the prerequisites common to all combination methods presented here are the following:
\begin{itemize}
\item all data have been fully calibrated (this would also include the possibility of self-calibration if applicable); 
\item astrometry should align; 
\item images have overlapping, well-sampled spatial frequencies, i.e. adequate coverage of the \emph{(u,v)}-plane \replyb{among combined datasets}; 
\item images have well-defined beams: primary beam (PB) of low-resolution image, $\text{PB}_{low-res}$ and primary beam of the high-resolution image, $\text{PB}_{hi-res}$; and 
\item bandwidth is narrow enough that the frequency dependent scaling of the baseline lengths and the primary beam can be ignored, although SDINT can in fact take this into account.  
\end{itemize}
We assume that flux scales are aligned, but further fine-tuning can be done for specific use cases.  Generally, every input image is assumed to be corrected for all instrumental effects including PB-response. We refer to the documentation of each method (i.e. CASA Guides, Github, etc.) for \replyb{guidance on handling units and} the specific treatments of the PB-response during the respective steps of the \replyb{data combination (sometimes referred to as DC)} process.  Finally, we use test cases without bright emission that extends or emerges beyond the area of the well-characterised PB-response.  The same methods should, in principle, be adaptable to other scenarios. }

\begin{deluxetable}{ccccccc}
\tablecaption{Summary of data combination methods\label{tab:methods}}      
\centering                                      
\tablehead{
\colhead{Methodology}
 & \colhead{Domain} & \colhead{Method} & \colhead{Task name} & \multicolumn{2}{c}{Input} & \colhead{Output}  \\
\cline{5-6}
\colhead{[1]} &\colhead{[2]}  &\colhead{}  & \colhead{} &\colhead{ interferometry} & \colhead{single dish} & \colhead{}  }
\startdata
\vspace{-0.4cm} & F/I & SDINT & {\tt sdintimaging} & vis & image & image  \\
\vspace{-0.2cm}
Before &  &  &  &  &  &   \\
& F & TP2VIS & {\tt tp2vis}[3] & vis & SD image  & pseudo-visibilities  \\
& & & {\tt tclean} & vis & pseudo-visibilities & image  \\
\cline{1-7}
During & F+I & MACF & {\tt tclean} & vis & image as model & image  \\
 &  &  & {\tt feather}  & image & image & image  \\
\cline{1-7} \vspace{-0.4cm}
 & F & Feather & {\tt feather} & image & image & image  \\
\vspace{-0.2cm}
After &  &  &  &  &  &    \\
 & I & FSSC & (script) & image & image & image
\enddata
\tablecomments{[1] indicates combination before, during, or after image deconvolution; [2] Fourier (``F'') or image (``I'') domain in which the method operates; [3] only available in CASA after import of the \texttt{tp2vis} package}
\end{deluxetable}

\subsection{Feather} \label{sec:feather} 

\subsubsection{Feather algorithm}

The Feather algorithm is probably the most well-known method for combining SD and interferometeric data. The algorithm is implemented as a CASA task of the same name. This method inverts the input images via Fourier transform (FT) and combines them in Fourier space (i.e. the \emph{(u,v)}-plane), weighting them according to the spatial frequencies of the response of each telescope. The combined data is then Fourier transformed back to the image plane. 


Although \texttt{feather} does regrid, in some cases, regridding may be better to do outside the task, for which case \texttt{imregrid} and \texttt{specsmooth} can be used. For a more comprehensive discussion of feather, see \citet{cotton2017}; the CASA documentation fully explains additional parameters.\footnote{An alternative implementation of the Feather algorithm is possible with the \texttt{uvcombine} package in Python, which implements a similar approach to CASA’s \texttt{feather} task, but with additional options. See \url{https://github.com/radio-astro-tools/uvcombine}.  \texttt{immerge} in Miriad offers the option \texttt{feather}, allowing the combination of two images with Gaussian PSFs (see \url{https://github.com/baobabyoo/almica} for implementation).  Additionally, J-comb \citep{Jiao2022} is a new linear combination technique that has been benchmarked against both CASA's \texttt{feather} and Miriad's \texttt{immerge} tasks, but is not further explored in this work.  See \citet{Jiao2022} for additional information about their investigation into the \texttt{feather} and \texttt{immerge} implementations.  }
 
 In more detail, the step-by-step order of operations undertaken with the \texttt{feather} task can be enumerated by the following:

\begin{enumerate}
    \item Regrids the SD/low-resolution image spatially and spectrally to the high-resolution image.
    \item FT both images, \reply{such that the next steps operate on the visibilities}.
    \item Scales the FT low-resolution image by the ratio of clean beams (CBs): 
        \begin{equation}\label{eq:clean_beams}
            CB_{hi-res}/CB_{low-res},
        \end{equation}
        \reply{in order to account for different beam-sizes.}
    \item Adds the FT of the high-resolution image $\times(1-\omega t)$ to the scaled FT-ed low-resolution image, where $\omega t$ is the FT of $CB_{low-res}$. \reply{In other words, the images have been weighted and combined in the $(u,v)$-plane.} 
    \item FT back to the imaging plane. 
\end{enumerate}

\subsubsection{Feather Parameters}
From the CASA documentation, the \texttt{feather} command inputs include those listed in Table \ref{tab:feather}.  Two particularly important parameters to consider for \texttt{feather} are \texttt{sdfactor} and \texttt{effdishdiam}. \texttt{sdfactor} is used to adjust the flux scale of the SD image.  \reply{It should be constrained by comparing flux where there is overlap between the spatial-frequencies in the low and high resolution images, and empirically, the value is most commonly set close to 1.0-1.2.} The parameter  \texttt{sdfactor} has parallels to \texttt{sdgain} in \texttt{sdintimaging} in SDINT (see \S~\ref{sec:sdintimaging}), but they are not identical mathematically. \texttt{effdishdiam} is the effective diameter of the SD telescope, corresponding to the ``low-res'' image in the \texttt{feather} command; the effective dish diameter depends on the aperture efficiency, and is generally determined/reported for a given telescope.  When Feathering, the weighting function for the SD data is usually the Fourier transform of the SD primary beam (PB$_{SD}$), but the weighting function can be altered by indicating a reduced SD diameter, corresponding to a different PB$_{SD}$. We did not further investigate the impact of \texttt{lowpassfiltersd}.

\begin{table}[!hbt]
\caption{Parameters for method Feather}\label{tab:feather} 
\label{table:2}      
\centering                                      
\begin{tabular}{llll}
\hline  \hline
   CASA Task & Parameter & Default & Description \\
\hline
  {\tt feather} &  imagename & = `' & name of output feathered image \\
    &highres & = `' & name of high resolution (interferometer) image \\
    &lowres &  = `' & name of low resolution (single dish) image \\
    &sdfactor & = 1.0 & scale factor to apply to SD image \\
    &effdishdiam & = -1.0 & effective diameter (in meters) of the SD telesope.  \\
    &lowpassfiltersd & = False & filter out the high spatial frequencies of the SD image\\
\hline

\end{tabular}
\end{table}

\subsection{Joint Deconvolution: SDINT} \label{sec:sdintimaging} 

\subsubsection{The SDINT algorithm}
The \emph{SDINT} algorithm \citep{rau2019} permits joint deconvolution of wideband SD and interferometer data, and has been implemented as the task \texttt{sdintimaging} in CASA\footnote{The task \texttt{sdintimaging} is incorporated in CASA 5.7/6.1 and subsequent releases.}.  We refer the reader to \citet{rau2019} for full details and explanatory diagrams, and we briefly describe the algorithm here.  The main inputs to the algorithm are the SD image and PSF (as a cube, if applicable), as well as the interferometric dataset.  \reply{Note that if no SD PSF is provided, \texttt{sdintimaging} can solve for the beam information based on the image header.}

The \texttt{sdintimaging} algorithm follows these steps: 
\begin{enumerate}
\item The process starts as for standard ``CLEANing'' with a ``major cycle" in which the interferometric data are gridded and Fourier-transformed into an image cube. At the same time, the  corresponding PSF is formed. 
\item Then the SD cube is combined with the interferometric cube in a Feathering step (see \S \ref{sec:feather}). 
\item A similar combination is applied to the interferometric PSF and the representative SD kernel to form a joint PSF.
The joint image and PSF cubes then form inputs to a standard CLEAN ``minor cycle" deconvolution. 
\item As the minor cycle limits are reached, model images from the deconvolution are translated back (depending on the chosen deconvolution algorithm) to model image cubes which are then (a) subtracted from the SD image cube to form a new residual SD image and (b) Fourier-transformed and degridded to form a new set of residual interferometric visibilities. 
\item These visibilities together with the residual SD image then form the input to the next major cycle and Feathering. For mosaicking, appropriate application of the primary beam correction takes place prior to deconvolution.
\end{enumerate}

Because it is based on the standard CLEAN family of algorithms, the task \texttt{sdintimaging} shares many parameters with the task \texttt{tclean} in CASA, as well as controls for a Feathering step to combine interferometer and SD datasets within the imaging iterations.
The most important parameter to control the Feathering step is \texttt{sdgain}. This parameter decides on the relative weight of interferometric and SD data similar to the relative weight for the visibilities from different baselines in pure interferometric imaging. The default setting for \texttt{sdgain} is 1.0, i.e. equal weight to both contributions. A value smaller than 1.0 will give the SD contribution less weight than the interferometric one, while a value greater than 1.0 gives it more weight. The flux scale of the joint image, however, will be kept constant. 
The best way to understand the role of \texttt{sdgain} is to think of the combination as forming a weighted mean of two measurements. In standard error propagation, the relative weight of different measurements of the same quantity should be 
\begin{equation}
    w_i = \frac{\sigma_i^{-2}}{\sum_j \sigma_j^{-2}}
\end{equation}
where $w_i$ is the weight and $\sigma_i$ is the error on the $i$th measurement. The natural \texttt{sdgain} value is thus a function of the noise or sensitivity of the SD and the interferometric contributions and should be close to 1.0 if the sensitivities are similar, larger than 1.0 if the SD observation is more sensitive and smaller than 1.0 if it is less sensitive than the interferometric one.

\subsubsection{SDINT Parameters}

\begin{table}[!hbt]
\caption{Parameters of method SDINT}\label{tab:paramsdint} 
\label{table:1}      
\centering
\begin{tabular}{llll}
\hline  \hline
    CASA Task & Parameter & Default & Description \\
\hline
  {\tt sdintimaging} &  usedata & `sdint' & output image type (int, sd, sdint) \\
    &vis & `' &  input interferometric visibility data \\
    &sdimage & `' & input single dish image \\
    &sdpsf & `' & input single dish PSF image \\
    &sdgain & 1.0 & A factor or gain to adjust single dish flux scale \\
    &dishdia & 100.0 & single dish diameter (in meters) \\
    &selectdata &  `' & enable data selection parameters, as in \texttt{tclean} \\
    \hline

\end{tabular}
\end{table}

Here we specify parameters specifically applicable when \texttt{usedata=`sdint'}, which is the case we explore for combination in this work.  \texttt{dishdia} should be considered as the effective dish diameter of the SD, and if {\tt sdpsf} is given as a dish diameter, this parameter will be ignored.  For a summary of pertinent parameters see Table \ref{tab:paramsdint}.

\subsection{Model-assisted CLEAN plus Feather method (MACF)} \label{sec:startmodel} 

There is another method which is in essence a variant of \texttt{feather}, and we refer to it as the Model-assisted CLEAN plus Feather method or MACF (for a list of the input parameters see Table \ref{tab:parammodel}).  It utilizes \texttt{feather} to ultimately combine the interferometric image and the SD image (as in \S \ref{sec:feather}), but takes a slightly different approach to generate the interferometric image used as input.  The parameter \texttt{startmodel} is used when invoking \texttt{tclean} on the interferometric data, providing the SD image (in units of Janskys per pixel) as the starting model image\footnote{A similar procedure to our MACF is documented by J. Kauffmann and is available at \url{http://tinyurl.com/zero-spacing}
}. 

The advantage of cleaning with the single dish image as a starting model is that the extended emission is conveyed in the model image, and then interferometer-based clean components are incorporated, before convolving with a clean beam for the final image restoration (adding residuals).  A caveat is that the zero-spacing flux, although initialized, is not constrained by \texttt{tclean}, so this should be considered an intermediate image only.  

The next step is to combine this interferometry-with-startmodel image with the SD image using \texttt{feather}.  This step in the combination procedure is identical to that described in \S \ref{sec:feather}. Examples of the application of the MACF method to observations can be found in \citet{dirienzo2015}, \citet{kauffmann2017} and \citet{hacar2018}.

\begin{table}[!hbt]
\caption{Parameters for method MACF (model-assisted CLEAN plus Feather)}\label{tab:parammodel} 
\label{table:3}      
\centering                                      
\begin{tabular}{llll}
\hline  \hline
   CASA Task & Parameter & Default & Description \\
\hline
   {\tt tclean} & startmodel & = `' & Name of low resolution (single dish) image  \\
   {\tt feather} & imagename & = `' & Name of output feathered image \\
    & highres & = `' & Name of high resolution (interferometer) image,\\ 
    &&&generated after using low resolution image as `startmodel' \\
    &lowres &  = `' & Name of low resolution (single dish) image \\
\hline

\end{tabular}
\end{table}

\subsection{TP2VIS} \label{sec:tp2vis} 

The package Total Power Map to Visibilities (TP2VIS)\footnote{The TP2VIS package is available via Github at: \url{https://github.com/tp2vis/distribute}} was developed for joint deconvolution of ALMA-12m, 7m, and TP data, and is thoroughly presented by \citet{koda2019}. We reiterate that Total Power, or TP, is another name used by radio astronomers for SD data. TP2VIS operates in CASA to convert a SD map into visibilities; these so-called ``pseudo-visibilities'' can then be used as input for joint deconvolution along with interferometric visibilities.  The technique has been presented and utilized by \citet[][\& references therein]{rodriguez2008,Kurono2009,pety2010,koda2011}, with the TP2VIS package being the first and only implementation into CASA. 

TP2VIS is roughly structured in four steps, with much more detail on each available in \citet{koda2019}:
\begin{enumerate}
\item The SD map is converted into the sky brightness distribution, to be observed as if by a virtual interferometer. 
\item The \texttt{tp2vis} function converts the brightness distribution into visibilities. 
\item Next, there is an option to set the weights of the TP visibilities according to a few weighting schemes, such as that they represent the rms noise of the original TP map or are set to a constant value.  
\item Finally, combination is done by deconvolving the pseudo-visibilities and the interferomentric visibilities jointly using the \texttt{tclean} task.
\end{enumerate}

The TP2VIS package also offers the function \texttt{tp2vistweak}, which attempts to fix the problem
of discrepant beamsizes of the dirty and clean beams in the image space after deconvolution, as detailed in \citet{jorsater1995}.

In our tests with CASA 6,\footnote{We note that TP2VIS was developed with an earlier version of CASA, as reported by \citet{koda2019}.} TP2VIS turned out to be less straight-forward to use effectively and consistently across different datasets, compared with the other combination methods described here. In order not to further delay the publication, we decided not to include it in the evaluation section of this paper. We have, however, included it in the accompanying software (\S \ref{sec:dc_script}).

\subsection{Faridani's Short Spacing Correction (FSSC) method}\label{sec:faridani} 

Implementations which utilize Fast Fourier Transforms (FFTs), such as the Feather method, may introduce artifacts in the final combined image due to structures that extend up to or beyond the edge of the image, creating a sharp discontinuity. This is the case for the \texttt{feather} task. \citet{faridani2018} introduced a combination method that works purely in the image domain to avoid the introduction of such artifacts; \reply{they provide a stand-alone Python code that we have also incorporated into our suite of combination methods in order for the implementation to be done consistently.} The method is succinctly summarized in Equation 6 of \citet{faridani2018}:
\begin{equation}
 \begin{aligned}
    I_{\rm missing} & = I^{\rm reg}_{\rm sd}-I^{\rm conv}_{\rm int} \\
    I_{\rm comb} & = I_{\rm int}+\alpha\cdot I_{\rm missing}. 
    \end{aligned}
\end{equation}

The missing flux ($I_{\rm missing}$) of the interferometry observations is determined by first taking the difference between the SD image at its native resolution $I^{\rm reg}_{\rm sd}$,  and the interferometer-only image that has been convolved and regridded to the angular resolution and pixel scale of the SD image, $I^{\rm conv}_{\rm int}$. The combined image is then the sum of the original interferometer-only image $I_{\rm int}$ and this missing flux image, also scaled by $\alpha$ which is the ratio of the clean interferometer beam to the SD beam to account for the difference in beam sizes (our Equation~\ref{eq:clean_beams}). In addition to mitigating potential FFT artifacts, this approach is also less computationally expensive since no FFTs are required. Similar to \texttt{feather}, an additional scaling factor can be applied to the SD image to compensate for differences in the flux scale beyond the differences in beam sizes. 

We refer to this process as the Faridani's Short Spacing Correction method (hereafter FSSC) and note that the output images are very similar to those from the Feather technique. While the concepts are equivalent, the FSSC method operates purely in the image plane, with no Fourier transforms, contrary to Feather.  This technique has successfully been used to combine images from the Large Millimeter Telescope, $\textit{Planck}$, and the Caltech Submillimeter Observatory to create a 1.1~mm image of the Central Molecular Zone of our Galaxy sensitive to all angular scales down to 10.5\arcsec\ \citep{tang2021a, tang2021b}.

\subsection{Software developed in this study}  \label{sec:dc_script}

In this work, we provide a single data combination script (called ``DC\_script'', that runs with Python and incorporates Astropy, \citet{astropy2018}) that performs (a) the data combination methods described in \S\ref{sec:methods} on simulated and observational datasets (\S\ref{sec:simdata} and \S\ref{sec:obsdata}) and (b) assessment methods described in \S\ref{sec:assess}. The script, data, and documentation are available at \packageurl. When running the script for the first time, follow the steps in the README file. Also available in the code repository are an overview of all steps, a "quick start" guide, and template parameter files. ``DC\_script'' is optimized to run non-interactively, but the user can choose to run interactively if needed in order to change certain parameters.  ``DC\_script'' also includes the full set of quality assessment metrics described in this work to evaluate the performance of the different data combination methods (\S~\ref{sec:assess}).


\section{Test datasets} \label{sec:testdata}
\subsection{Simulated continuum observations} \label{sec:simdata}

\begin{figure}[hbt]
\centering
\includegraphics[width=0.6\textwidth]{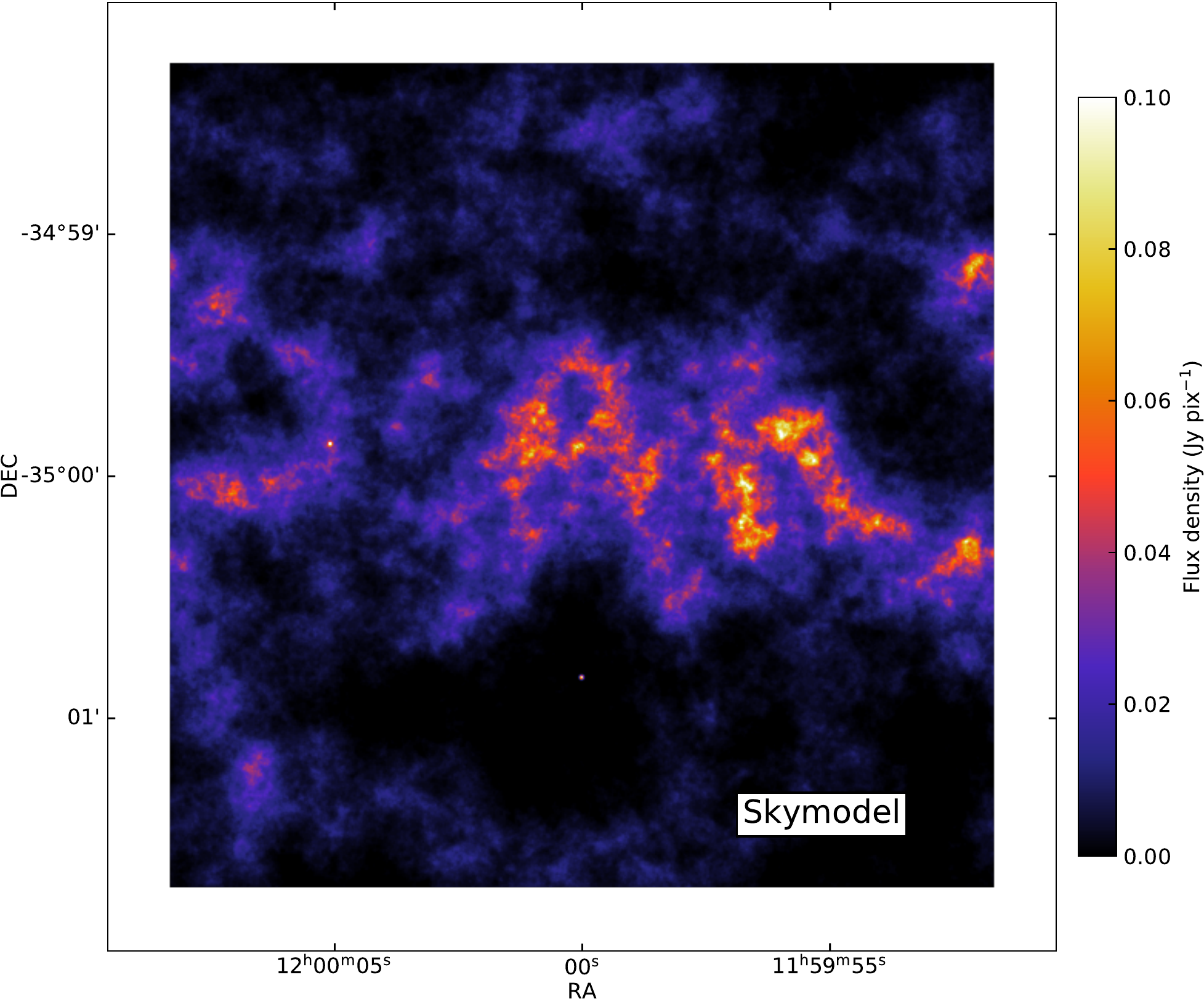}
\caption{The synthetic "skymodel" image used as input for the simulated observations.}
\label{fig:skymodel}
\end{figure}

Only with simulated observations can the accuracy of the image reconstruction be properly measured, since in that case the ``true'' intensity distribution on the sky is known. For efficiency we try to capture a large range of possible source structures in a single test image to study. An elegant way of achieving this is to emulate a multi-scale molecular cloud with a power spectrum that follows a power law. We follow the procedure described in \citet{koda2019}. In addition we insert two bright point sources, one in a bright and one in a faint region. We call the resulting model the ``skymodel".
The model has $4,096^2$ pixels with a size of 0.05\arcsec\ each, resulting in an image size of 204.8\arcsec\ each side, shown in Fig. \ref{fig:skymodel}.  

With the ``skymodel'' as input, we ran the {\tt simobserve} task in CASA version 5.6.1 to create simulated observational data in MeasurementSet format (MSs). Following the ALMA Technical Handbook \citep{remijan2019}, we adopted two configurations for 12m array (C-4 and C-1 in cycle 7) along with ACA 7m and SD. The expected beam size for this configuration is $\sim$1\arcsec\ for an observing frequency of 115 GHz, and we covered 2 GHz bandwidth. The center coordinate of the field was set as (RA,Dec) $=$ (12h00m00.0s, $-35$d00m0.0s). For 12m (7m) array, there are 52 (17) fields to cover the simulated image and each field is set to be visited 30 (90) times per \reply{observing session on different days} with the integration time of 10 seconds per field.  The SD observation was set up with 169 fields, each visited 9 times per day. Consequently, the total integration time to cover the skymodel image with 12m/7m/SD array is 1040/1020/3549 minutes, respectively. Both configurations of the ALMA-12m array have the same integration time. \reply{This particular setup is to simulate a realistic ALMA observation and optimise \emph{(u,v)} coverage.}  To ensure ideal \emph{(u,v)} coverage even for the 7m array at each field,  observations for 12m and 7m arrays are set to be carried out over four separate days with a different hour angle at the start of observations, which is $-2$, $-1$, 0, and 1$h$. For the SD array, observations are set to be carried out over 14 days, due to the large number of fields. The achieved beam sizes are 57\arcsec\ for the SD data and 1.2\arcsec\ for the interferometric data. Relevant parameters for the simulated observations are summarized in Table \ref{table:data}.

\begin{figure}[hbt]
\centering
\includegraphics[width=1.0\textwidth]{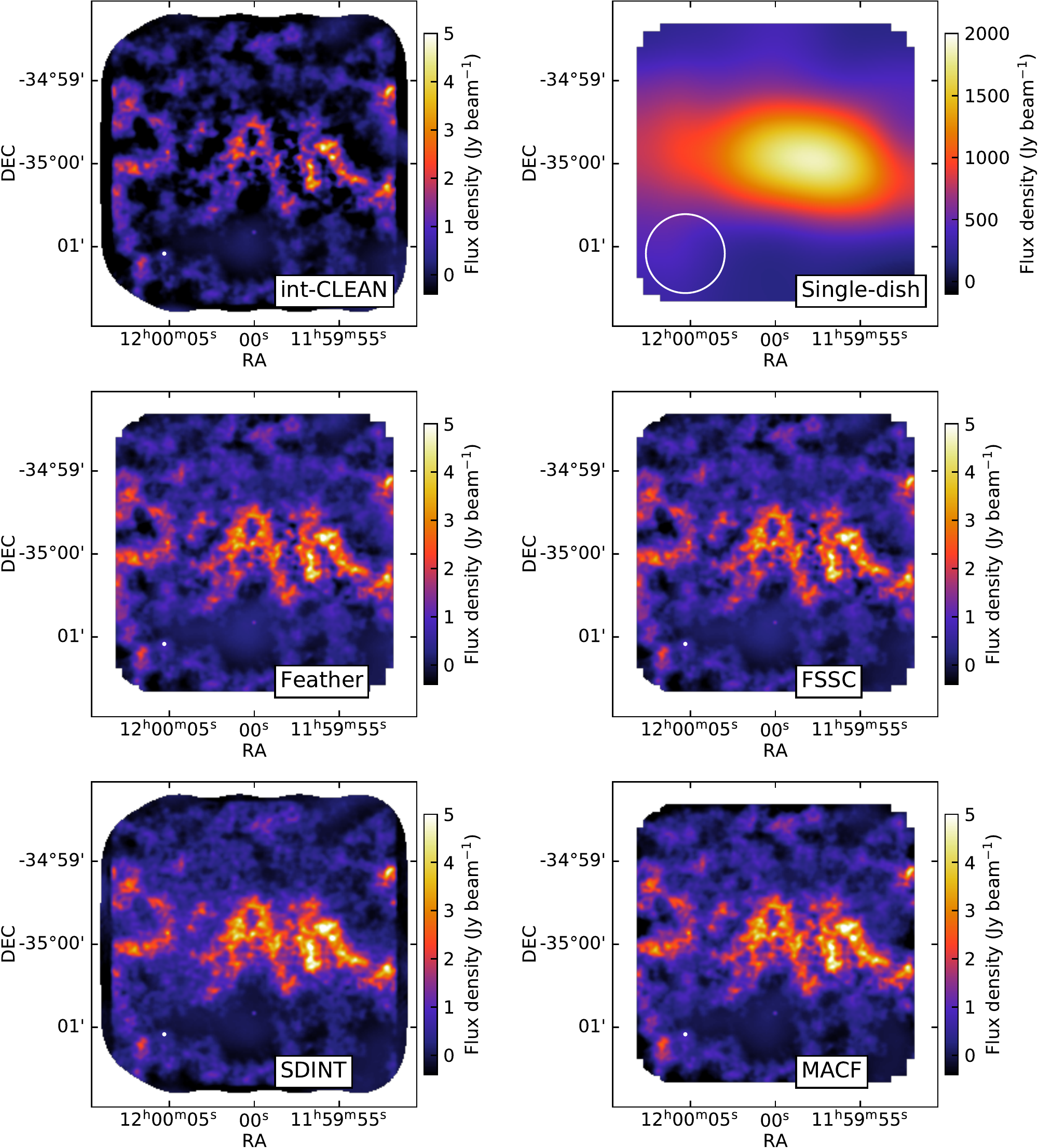}
\caption{Maps of the simulated observations based on the skymodel image in Fig.~\ref{fig:skymodel} obtained with different combination methods. From left to right and from top to bottom:  int-CLEAN (interferometric-only), SD-only, Feather, FSSC, SDINT, and MACF. All interferometric and combined images are convolved into a final circular beam of 1.7\arcsec, and are displayed within the same colour scale to facilitate their comparison. The beamsize is indicated in the lower left corner of each image.
\label{fig:skymodelmaps}}
\end{figure}

Figure \ref{fig:skymodelmaps} shows the images obtained from the simulated skymodel observations with interferometry-only (hereafter int-CLEAN), the SD data only, and with the different data combination methods (Feather, FSSC, SDINT, and MACF) described in \S \ref{sec:methods} using both the simulated interferometric and the simulated SD data.

\subsection{Observational, spectrally resolved data} \label{sec:obsdata}

Working with real observational data, one can only make assumptions about the true sky brightness distribution. However, one condition which the combined interferometric and SD image must at least fulfill is that it recovers most of the flux visible in the SD image because in that image, no filtering of spatial frequencies has taken place.

Furthermore, as mentioned in the introduction, there is a danger of the deformation of spectral features when no data combination is applied. So, the flux-recovery condition needs to be tested for each spectral channel separately and the results have to be compared between channels.  Considering this, we therefore choose the spectrally resolved observation of the galaxy M100 as an example of real observational data.

The galaxy M100 (NGC 4321) was observed by ALMA as Science Verification (SV)\footnote{\url{https://almascience.org/alma-data/science-verification}} data, and has been featured in the {\it CASA Guides} series as an example of data combination using the Feather method \footnote{\url{https://casaguides.nrao.edu/index.php/M100_Band3}}. The `grand-design' barred spiral morphology of this galaxy, seen relatively face-on, reveals structures with different physical size scales.  The data cube that we present here reveals in 50 channels the CO ($1-0$) (115.271 GHz, ALMA Band 3) molecular emission. On each side of the spectral line, additional 10~channels are included in the cube to show the spectral baseline.  Observations were made of a 47 pointing mosaic with the 12m Array, a 23 pointing mosaic with the 7m Array, and an on-the-fly map with the SD array, effectively recovering a range of scales down to the 12m Array resolution of $<2\arcsec$.   

\begin{figure}[hbt]
\centering
\includegraphics[width=0.95\textwidth]{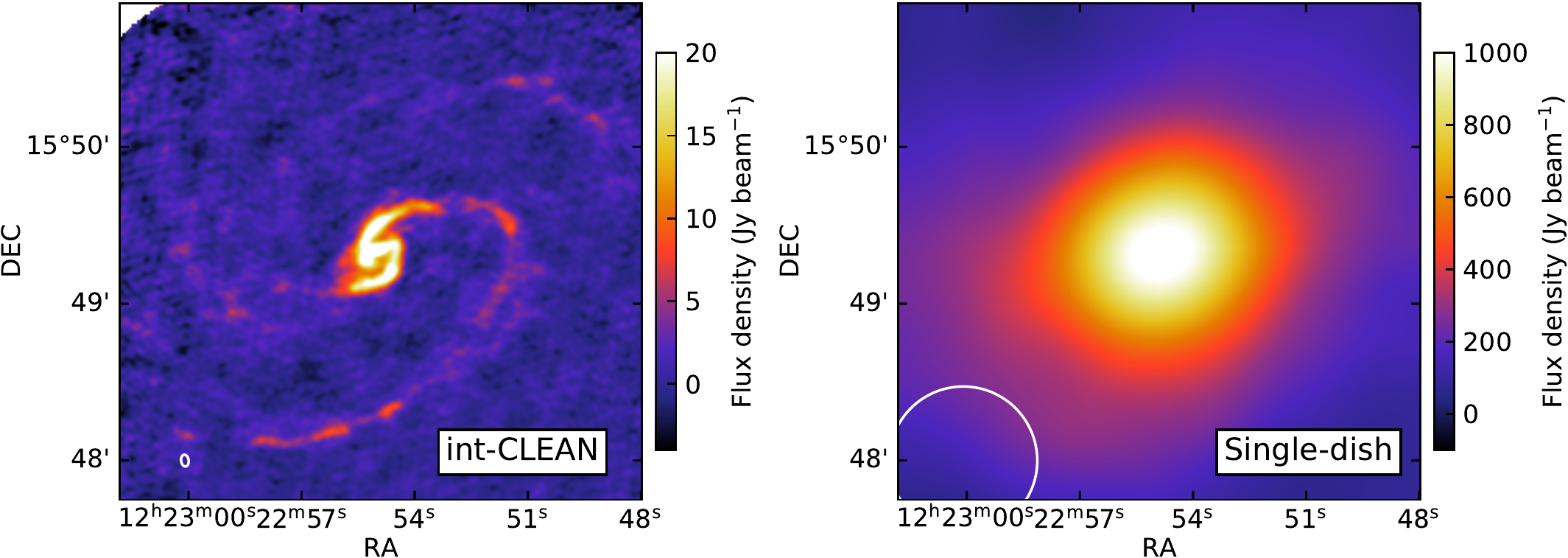}
\caption{Integrated emission (moment 0, over all 70 channels) maps of our M100 data comparing the interferometry-only (left) and the SD image (right). The circles in the lower left represent the respective angular resolution (beam size).}
\label{fig:m100maps}
\end{figure}


This dataset allows us to showcase the necessary preparatory steps of header manipulation, axis reordering, and regridding. The observational details are given in Table \ref{table:data}. 
Figure \ref{fig:m100maps} shows the interferometric and the SD M100 images (in each case the integrated emission over all 70 channels).


\begin{table}[!hbt]
\caption{Details of the datasets}\label{fig:aboutdata} 
\label{table:data}      
\centering                                      
\begin{tabular}{l c c}          
\hline\hline                        
   Parameter& Skymodel & M100 \\
\hline
Phase center RA  & 12h00m00.0s & 12h22m54.900s \\      
Phase center DEC  & -35d00m0.0s & +15d49m15.000s\\
Rest frequency  & 115GHz  & 115.27120GHz \\
$V_{LSR}$ &\nodata& 1575 km/s  \\
$\Delta V$ (channel width)  &\nodata& 5 km/s  \\
Velocity range imaged  &\nodata& [1400-1745] km/s  \\
Map size  & 204\farcs8 $\times$ 204\farcs8 &400\arcsec $\times$ 400\arcsec \\
12m-array pointings  & 52 & 47 \\
7m-array pointings  & 17 & 23\\
12m-array beamsize & 1\farcs2 $\times$ 1\farcs0 (PA=-84\fdg8)&  4\farcs0 $\times$ 2\farcs6 (PA=-88\fdg2)\\
Range of 12m baselines  & [15 - 784] m &[15 - 200] m\\
Range of 7m baselines  & [9 - 45] m&[9 - 42] m\\
\hline                                             
\end{tabular}
\end{table}


\section{Assessment metrics} \label{sec:assess}

Assessing the quality of the output image of a data combination is of utmost importance in the application and interpretation of the different methods \citep[e.g., see][]{Pety2013}. The intrinsic definition of quality refers to the accuracy in which the combined data reproduce the true sky emission. In empirical research, the definition of an absolute quality assessment is hampered by the lack of a-priori expectations for this true sky distribution.
Instead, the quality assessment of a given target image $I_\nu(x,y)$ can be evaluated against the emission constraints present in a reference image $R_\nu(x,y)$. Here $I_\nu(x,y)$ is not necessarily identical to the data combination output image but can be a modified version of it, e.g. a version of this image convolved to the SD resolution. 

This approach is motivated by our observational products. In real observations the emission of any interferometric or data combination map can be compared against its SD counterpart in order to evaluate the total flux and emission structure recovered at the SD resolution. Following this same approach, synthetic interferometric data and combinations can also be evaluated against simulations at the maximum resolution provided by the interferometer once both are regridded into a common spatial and spectral coordinate system. Our analysis examines both the synthetic data from \S \ref{sec:simdata} and the real observations from \S \ref{sec:obsdata}.

We define as reference image $R_\nu(x,y)$ either the SD image or the synthetic model used for comparison, while the target image $I_\nu(x,y)$ is assumed to be the combination result image convolved to the resolution of the reference image. For spectrally resolved data (cubes) our quality assessments will also detect variations of the quality with frequency, i.e. between channels.

We note that the effective comparison of the two images, $R_\nu(x,y)$ and $I_\nu(x,y)$, must be performed pixel-by-pixel at the same angular resolution (typically corresponding to one defined by $R_\nu(x,y)$, i.e. SD or synthetic model convolved to low-resolution). \reply{Thus, if $I_\nu(x,y)|_0$ is the image at {\it native} interferometric
resolution or the data-combined image, then $I_\nu(x,y)$ refers to the image convolved and (spatially and spectrally) regridded to the reference image parameters.}

The spatial frequency filtering which occurs in interferometric observations
usually affects the structure, spectral distribution, and absolute flux of the recovered emission. Assessing the quality of a given dataset therefore requires considering fundamental parameters such as the total recovered flux and its dependence on scale and intensity, as well as the variation of these indicators across the target field. These analyses should be carried out in statistical terms in both maps and cubes in order to make the assessment objective. In the following sections we introduce a new set of image quality assessment metrics for the analysis of interferometric observations and their combination between multiple arrays and SD data, and illustrate their use on the comparison of int-CLEAN and Feather method results. Feather is used as a benchmark of any combination technique presented here, since it is the most common method used in CASA.  Comparison of combination methods will be done in a following section.

\subsection{Assessment mask: definition and adaptive thresholding}\label{sec:AM}

A detailed analysis of the image quality requires a careful definition of the so-called assessment region, that is the sub-sample of (x,y) pixels (or (x,y,$\nu$) voxels for cubes) in which our assessment metrics will be evaluated. Not necessarily all pixels in a map are relevant for the assessment of recovered flux. Regions devoid or showing low levels of emission should be \reply{used specifically for determining the RMS noise of an image, as they are dominated by pure instrumental noise, and otherwise excluded (masked) from the image quality assessment}. Ideally, any quality assessment should also consider the structure of the astronomical object as well as include the instrumental response. The definition of assessment areas (e.g. rectangular region) or global intensity thresholds can simplify this analysis. To improve these basic -- and sometimes arbitrarily defined -- criteria, here we introduce a more general assessment mask $AM_\nu(x,y)$ depending on the properties of the corresponding data products.

The definition of $AM_\nu(x,y)$ considers three conversion factors depending on the beam sizes $\theta_0$ and $\theta_R$ of the (native) $I_\nu(x,y)|_0$ and $R_\nu(x,y)$ images, respectively, as well as the primary beam map $PB_\nu(x,y)$. The mask is determined according to the following steps:
\begin{itemize}
\item We define the ``mask threshold'' at a level of 
$3\times RMS_{mask}$, where $RMS_{mask}$ is the off-source noise level of the target \reply{(in this case, interferometric)} image at its native resolution $\theta_0$.
\item This single RMS value is corrected for the effective antenna response in each pixel defined by the primary-beam correction (PB) per pixel, that is, $1/PB_\nu(x,y)$.  The response per-pixel is therefore $RMS_{mask}/PB_\nu(x,y)$.
\item Next, the mask threshold needs to be evaluated at the reference resolution $\theta_R$ for our assessments and therefore be corrected by a factor $\propto 1/\sqrt{N}$ where $N$ is the number of $\theta_0$ beams within $\theta_R$, that is, $\sqrt{\theta_0^2 / \theta_R^2} = \theta_0 / \theta_R$. 
\item Finally, the resulting mask threshold value, originally in units of flux per $\theta_0$ beam, needs to be converted into $\theta_R$ units ($\propto 1/\theta_R^2$)  requiring an additional $\theta_R^2 / \theta_0^2$ factor.
\end{itemize}

Altogether, we thus define our assessment mask $AM_\nu(x,y)$ as:
\begin{equation} \label{eq:mask_th}
    AM_\nu(x,y) = \left( 3\times RMS_{mask} \right) \cdot \frac{1}{PB_\nu(x,y)} \cdot \frac{\theta_R}{\theta_0},
\end{equation}
\reply{where the $\nu$ sub-script denotes that the mask can be applied to a single image, or plane of a cube.}  We apply the same $AM_\nu(x,y)$ mask to both $R_\nu(x,y)$ and $I_\nu(x,y)$ images, respectively, blanking all corresponding pixels where emission $R_\nu(x,y)< AM_\nu(x,y)$ at the reference resolution; \reply{in other words, the so-called ``mask'' retains the values where the assessment should apply}. Compared to a single intensity threshold, $AM_\nu(x,y)$ effectively applies an {\it adaptive thresholding} taking into account the different image resolutions and the spatial instrumental response, \reply{as well as channel-by-channel variations in emission structure (in the case of cubes), where the same mask is applied to all respective images.}

\subsection{Accuracy parameter \& Fidelity: assessing flux recovery}\label{sec:apar_fid}
The assessment of recovered flux has been quantified in several distinct but related ways in the literature since radio interferometers became available. A commonly used parameter is image "fidelity" \citep[see, e.g. ][but a detailed definition follows below]{cornwell1993}. This quantity is unsigned (i.e. doesn't indicate if too little or too much flux is recovered) and has no optimal value (its best value is ``as large as possible").  In this work, we introduce a new assessment parameter which we call the {\it Accuracy parameter} (A-par). A-par is defined as:
\begin{equation} \label{eq:Anu}
    \begin{split}
    A_\nu(x,y)&=\frac{I_\nu(x,y)-R_\nu(x,y)}{|R_\nu(x,y)|},\; \; \;  -\infty < A_\nu(x,y) < \infty
    \end{split}
\end{equation}

A-par represents the relative difference between the flux of the input image $I_\nu(x,y)$ and reference image $R_\nu(x,y)$. Values of A-par range between negative and positive infinity. The ideal value is zero. A-par is the signed relative error of the recovered emission after combination with respect to the reference image, where its sign indicates whether the emission is underestimated ($A_\nu<$~0) or overestimated ($A_\nu >$~0). A perfect matching between the input $I_\nu(x,y)$ and reference $R_\nu(x,y)$ images thus corresponds to minimizing A-par in all pixels. 

A-par can be understood as the approximate inverse of the {\it Fidelity} parameter ($F_\nu$), such that $A_\nu(x,y) \approx F_\nu(x,y)^{-1}$.  Note that Fidelity is defined slightly differently by different authors. A compilation of the different conventions for ALMA, VLA, and ngVLA, can be found in \citet{mason2021}. We adopt the following definition:
\begin{equation} \label{eq:Fnu}
    F_\nu(x,y)=\left|\frac{R_\nu(x,y)}{I_\nu(x,y)-R_\nu(x,y)} \right|, \; \; \;  0 < F_\nu(x,y) < \infty
\end{equation}
When used for image quality assessment, higher Fidelity values indicate stronger similarities between the input $I_\nu(x,y)$ and reference $R_\nu(x,y)$ images (if $I_\nu(x,y)\sim R_\nu(x,y)$ then  $F_\nu\gg 0$). Fidelity becomes a powerful assessment method for the relative flux comparison of distinct input images (e.g. $F_{\nu,1} > F_{\nu,2}$).
However, as mentioned above, the target value of this Fidelity parameter is by construction undefined (i.e. if $I_\nu(x,y)\rightarrow R_\nu(x,y)$ then  $F_\nu\rightarrow\infty$), which limits its use during the assessment of individual images in absolute terms. In comparison, A-par provides additional information about the flux recovered (i.e. considering both A-par value and sign), and therefore we consider it more informative. Still, our analysis incorporates both A-par and Fidelity assessments for a full description.

\begin{figure}[t]
\centering
\includegraphics[width=0.9\textwidth]{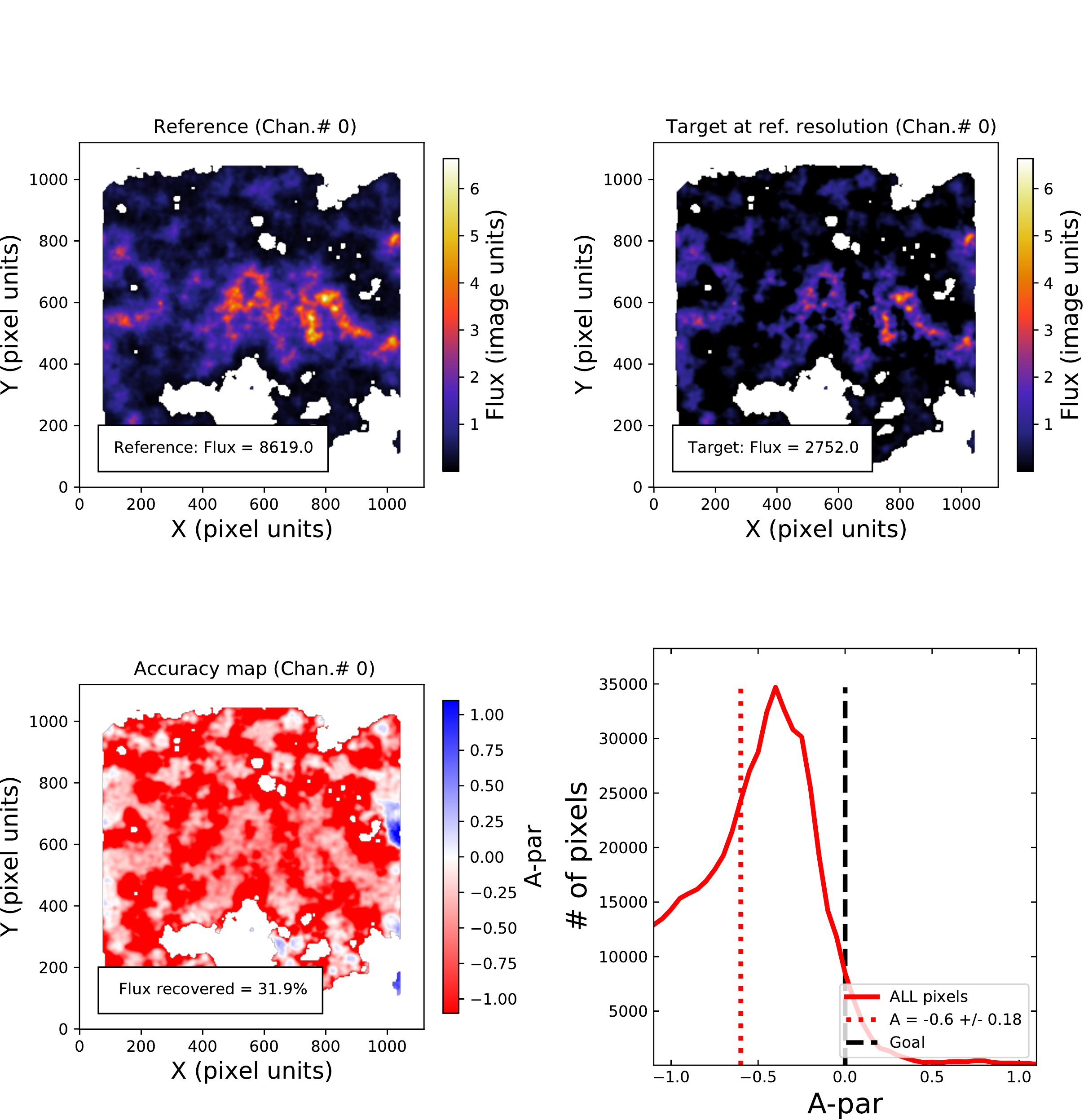}

\caption{Accuracy parameter (A-par) assessment for the int-CLEAN reduction of our synthetic continuum skymodel data (\S \ref{sec:simdata}): 
{\bf (upper left panel)} skymodel used as reference image $R_\nu(x,y)$,
{\bf (upper right panel)} feather image used as target $I_\nu(x,y)$,
{\bf (lower left panel)} A-par map,
and {\bf (lower right panel)} A-par histogram. Note that all results are obtained at a common resolution of 2.0\arcsec\ and display all pixels within the assessment mask $AM_\nu(x,y)$ (in color). Total flux values for skymodel (reference; top right) and int-CLEAN (target; top left) images, as well as the amount of flux recovered (bottom left) are indicated in the legends. 
The histogram also includes the mean and standard deviation values for the observed A-par distribution. The vertical dashed lines show the mean A-par (red) and its desired value zero (black).
\label{fig:skymodelb_apar_map_intclean}}
\end{figure}

\begin{figure}[t]
\centering
\includegraphics[width=0.9\textwidth]{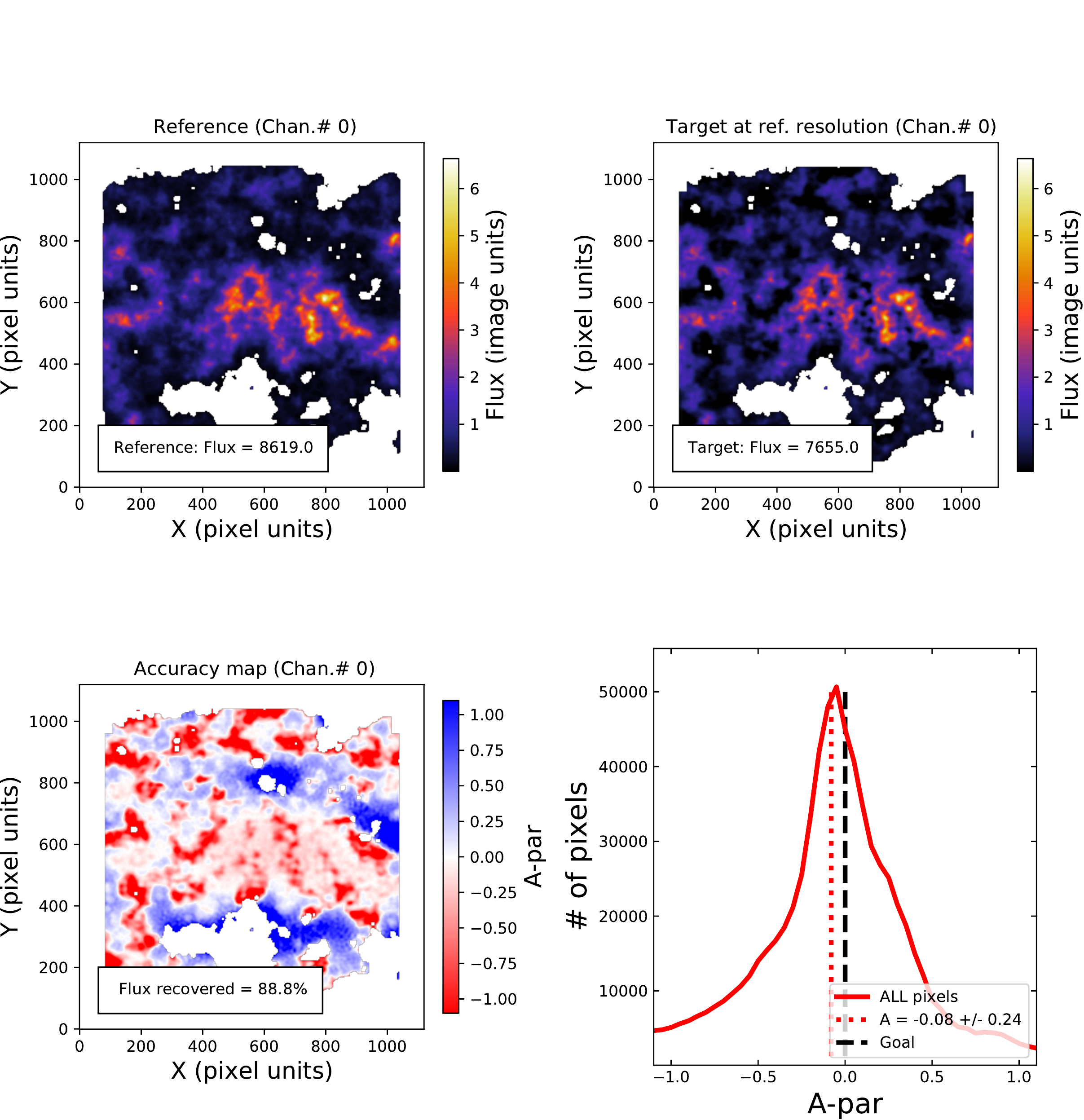}

\caption{Accuracy parameter (A-par) assessment for the Feather of our synthetic skymodel data (see also Fig.~\ref{fig:skymodelb_apar_map_intclean}).
\label{fig:skymodelb_apar_map}}
\end{figure}

We calculate both A-par and Fidelity parameters pixel-by-pixel (or voxel in cubes) in our images, that is, $A_\nu(x,y)$ and $F_\nu(x,y)$, for all pixels passing the assessment mask $AM_\nu(x,y)$ (see \S\ref{sec:AM}). We carry out these calculations by convolving and regridding both $I_\nu(x,y)$ and $R_\nu(x,y)$ images into a common beam $\theta$ and grid (x,y), respectively, typically corresponding to those of the reference image $R_\nu(x,y)$. 
Additionally, we prefer to make assessments per-pixel (Eqs. \ref{eq:Anu} and \ref{eq:Fnu}), in order to portray the particular regions of the image that are more or less accurate, rather than effectively averaging over an entire image.

\begin{figure}[t]
\centering
\includegraphics[width=0.9\textwidth]{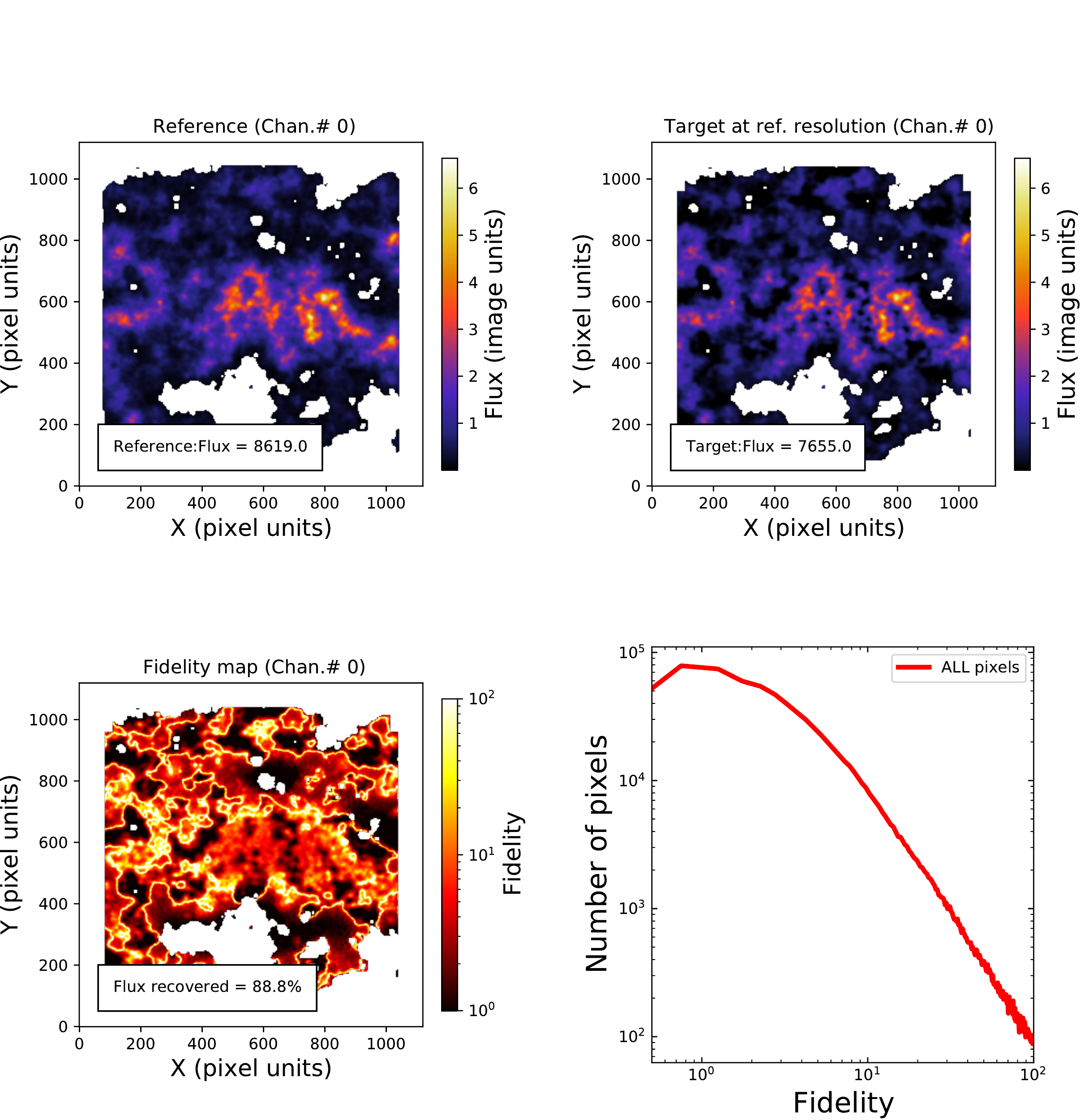}
\caption{ \label{fig:skymodelb_fidelity_map} Fidelity assessment of using Feather on our Skymodel data (\S \ref{sec:simdata}). Panels are analogous to those for A-par in Fig.~\ref{fig:skymodelb_apar_map} except that here a mean value is not plotted since the shape of the distribution is not well characterized by it. An ideal value does not exist by definition.}
\end{figure}

We show the A-par maps for the assessment of the data combination of our synthetic Skymodel data (\S \ref{sec:simdata}) using int-CLEAN and Feather (\S \ref{sec:feather}) in Figures \ref{fig:skymodelb_apar_map_intclean} and \ref{fig:skymodelb_apar_map}, respectively.
In both figures we display the reference (i.e. the true Skymodel; upper left panel) and target (Feather image; upper right panel) images, both convolved into a circular beam of 2\arcsec. An assessment mask as described by Eq. \ref{eq:mask_th} has been applied. The masked pixels are shown with white in the upper two panels of the figure. The lower panels show the A-par map (lower left panel) and histogram obtained from the pixel values of the A-par map (lower right panel).  
Figure \ref{fig:skymodelb_fidelity_map} shows similar plots but for the Fidelity parameter. 
This representation allows us to intuitively identify variations of flux recovered in different regions of our maps (A-par or Fidelity maps) as well as to statistically describe these properties over the target field (histograms).

\begin{figure}[t]
\includegraphics[width=0.5\textwidth]{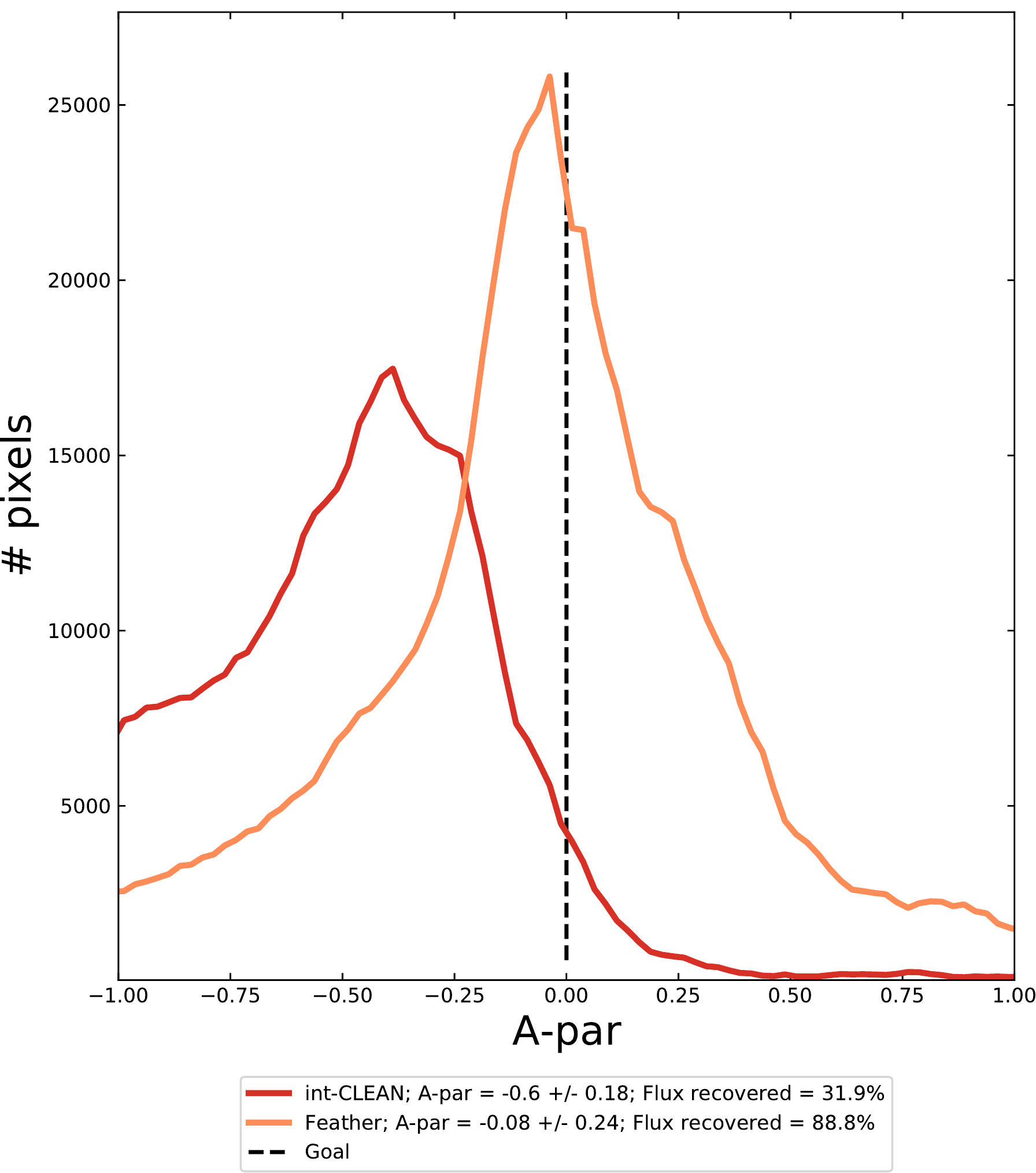}
\includegraphics[width=0.5\textwidth]{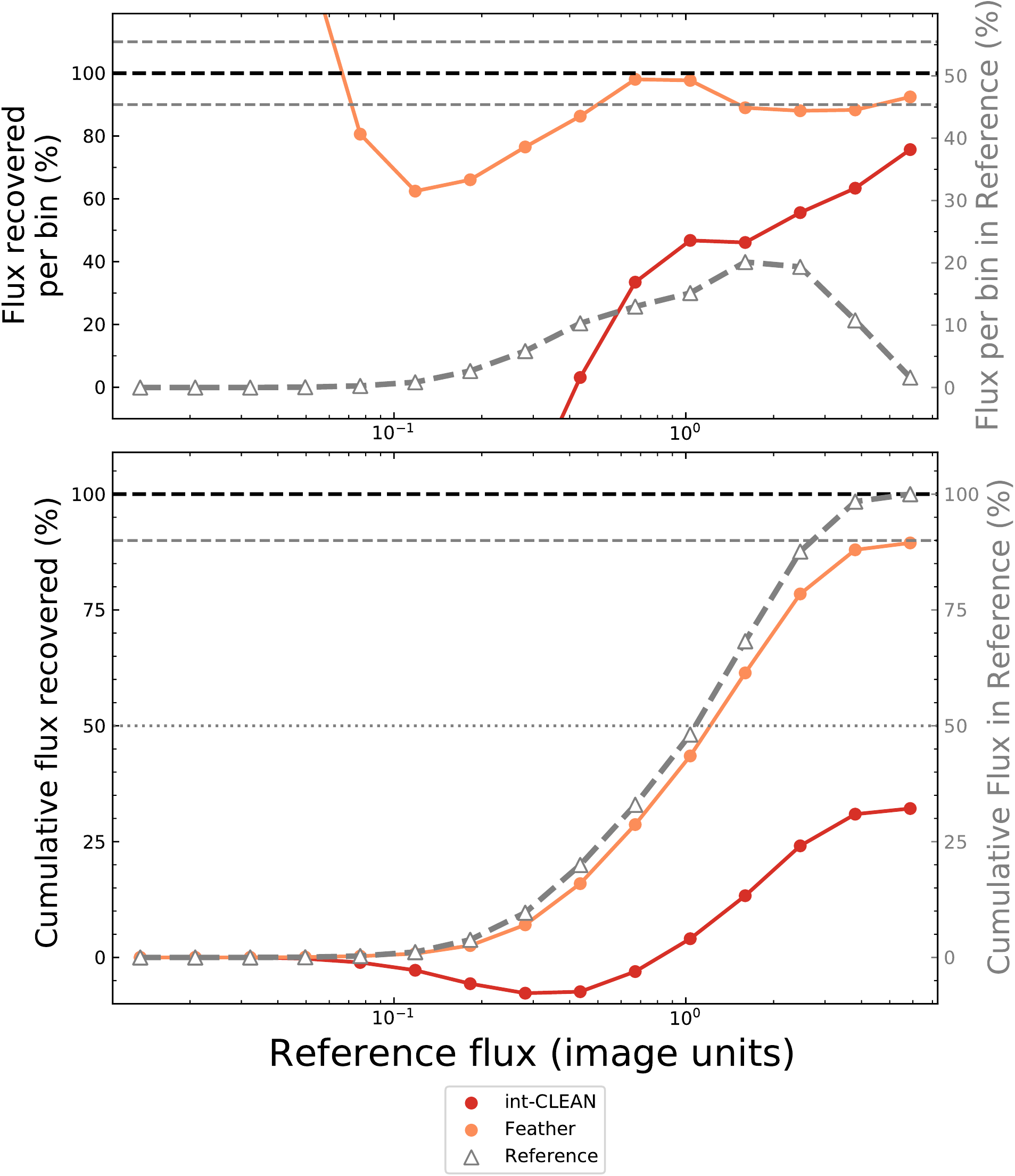}
\caption{Comparison of the A-par distributions for the interferometer-only CLEAN (red) and Feather-combined (orange) results on our skymodel data.
{\bf (Left panel)} Histogram of all the A-par values within our assessment area, where a perfectly recovered pixel would contribute a value of 0. 
{\bf (Right panel)} Flux recovered per flux bin shown in individual bins (upper sub-panel) and cumulative (lower sub-panel). In both plots, the corresponding flux of the reference image, i.e. the expected recoverable flux, is shown in grey and corresponds to the right axis, also in gray. In the upper right panel, the flux recovered per bin is shown in percentage with respect to the total flux in the image (see right vertical axis). Horizontal dashed lines, with guiding values of 100$\pm$10\% in the upper sub-panel and 50, 90, and 100\% in the lower sub-panel, respectively, are included to improve the readability of the plot.
\label{fig:skymodelb_apar_all}}
\end{figure}

The analysis of the histograms and distribution of the A-par values in our images allows us to quantify the quality of a data combination technique (see Figs.~\ref{fig:skymodelb_apar_map_intclean} \& \ref{fig:skymodelb_apar_map}, lower right panel). The mean value and dispersion of such distributions provides information about the typical fraction of recovered flux per pixel. High quality data combination techniques would produce A-par histograms with mean values close to zero ($A=0$) and a small width ($\sigma(A)\rightarrow 0$) showing a narrow width distribution that is approximately Gaussian. 

Similar comparisons using Fidelity would produce distributions with increasingly high peaks and more pronounced skewness towards large Fidelity values. In both cases the inspection of the A-par and Fidelity maps (in Figs.~\ref{fig:skymodelb_apar_map_intclean}, \ref{fig:skymodelb_apar_map}, and \ref{fig:skymodelb_fidelity_map}) can be used to identify spatial variations and systematic effects on the quality of the data.

Multiple histograms can be overplotted in order to compare different images obtained from the same observation. We show an example of such a comparison in Figure~\ref{fig:skymodelb_apar_all} (left panel) which illustrates the very significant improvement on the recovery of the true sky emission produced by applying a data combination technique (here Feather, blue histogram; A-par$\approx -0.08\pm 0.24$) in comparison to using the interferometric-alone int-CLEAN reduction (red histogram; A-par$\approx-0.60\pm0.43$). These improvements translate into the total amount of flux recovered in each case. While int-CLEAN only recovers $\sim$~32\% of the total reference flux, Feather manages to improve these values up to 89\%.

\subsection{A-par vs flux: flux recovery as a function of intensity} 

\begin{figure}[tb]
\includegraphics[width=0.5\textwidth]{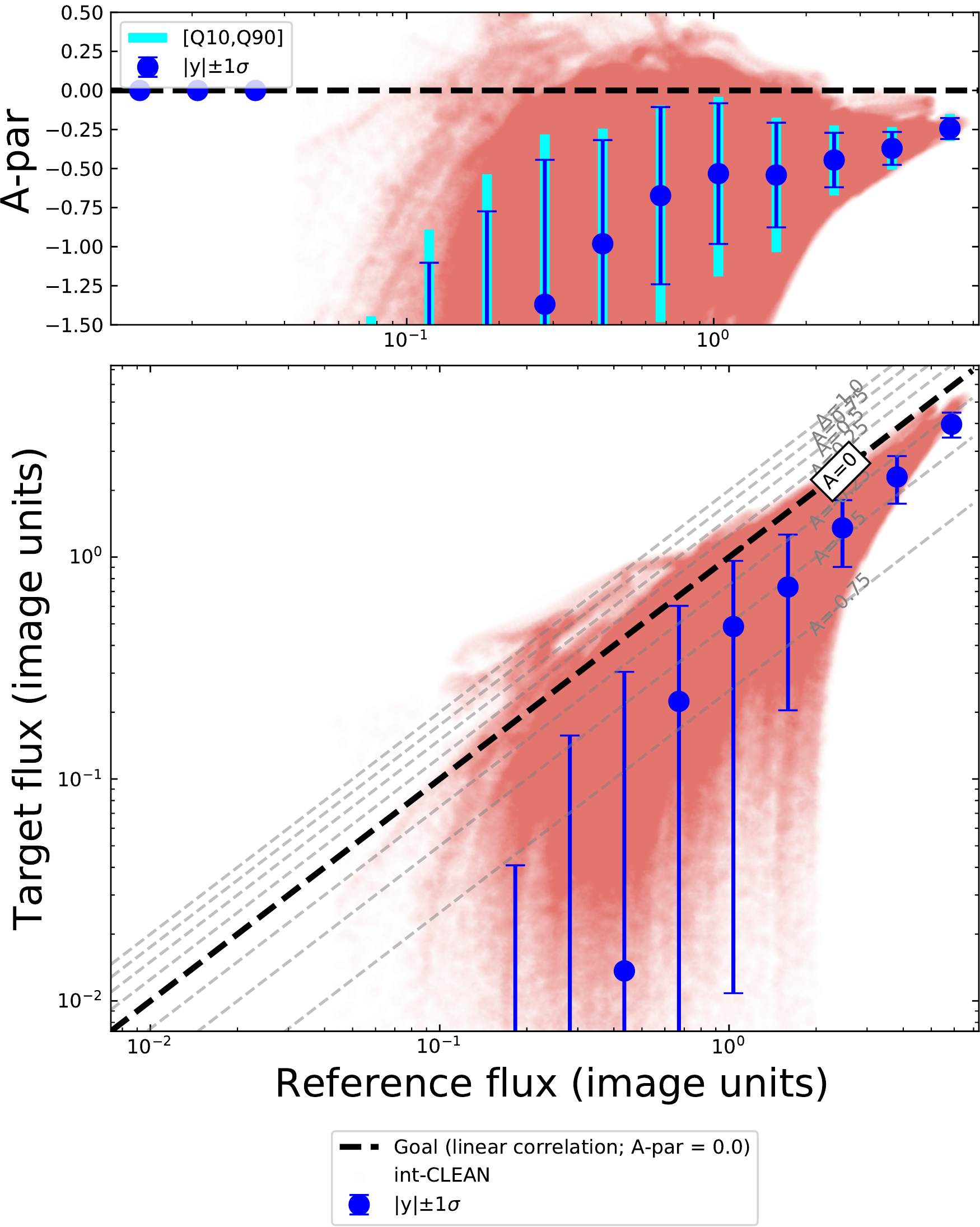}
\includegraphics[width=0.5\textwidth]{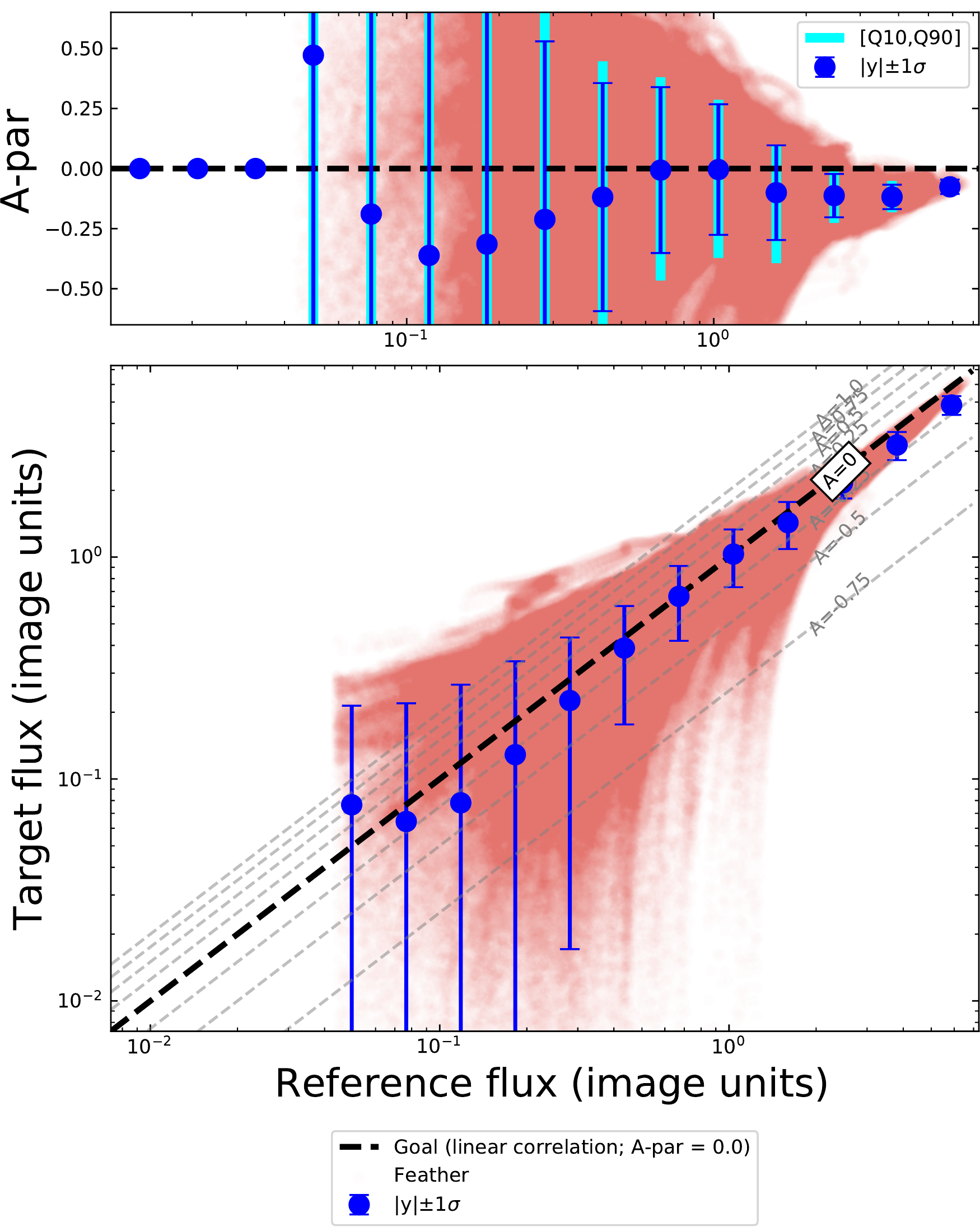}
\caption{A-par values as function of intensity for both int-CLEAN, i.e. pure interferometric data, (left) and Feather (right) methods for our skymodel synthetic data. Lower panels: pixel-by-pixel  comparison between the reference flux $R_\nu(x,y)$ and recovered flux in the target image $I_\nu(x,y)$ for pixels included in the assessment mask $AM_\nu(x,y)$ (red points). The dashed lines indicate the expected correlations for different A-par values. 
Upper panels: Individual A-par $A_\nu(x,y)$ values as function of the reference flux $R_\nu(x,y)$. We represent the mean and standard deviation in 10 intensity bins in dark blue. Also, we display the interquartile [Q10,Q90] range in cyan.
In both lower and upper panels, we indicate the expected distribution limits for the case of perfectly recovered emission (i.e. $A=0$) showing only white noise deviations with $\sigma=0.1$ (in image units).   \label{fig:skymodelb_aparveflux}}
\end{figure}

The distributions of A-par and Fidelity parameters provide information about the global properties of the flux recovery in our data. However, the fraction of recovered flux in the target image \reply{depends on the sky brightness temperature}. This is particularly true if the target field exhibits large regions with low-intensity but extended flux in comparison to bright and compact emission spots  (see variations of the corresponding A-Par and Fidelity maps in lower left panels of Fig.~\ref{fig:skymodelb_apar_map} and Fig.~\ref{fig:skymodelb_fidelity_map}, respectively). In these cases it is instructive to investigate not only the total A-par distribution (\S\ref{sec:apar_fid}) but also the variation of this parameter as a function of the reference flux. 

We investigate the dependence of the flux recovery as a function of flux in Fig.~\ref{fig:skymodelb_apar_all} (right panel). We represent both the fraction (percentage) of the flux recovered in individual (upper sub-panel) and cumulative (lower sub-panel) diagrams. Critical for their interpretation, we overplot the contribution of the reference flux (again per flux bin and cumulative, respectively) in both sub-panels. Maximizing the flux recovered is key for any data combination method, and most significant in the case of flux bins with large contributions to the total image flux. In our Skymodel example, bins with intensities between $\sim$~0.4 and 4.0 (in image units) each contribute to $>$10\% of the total flux of the reference image (see grey distribution in upper sub-panel, with reference to the right y-axis also in grey) becoming the most relevant flux bins.

In Figure~\ref{fig:skymodelb_aparveflux} we show a direct pixel-by-pixel comparison between reference flux $R_\nu(x,y)$ in the Skymodel and the recovered flux $I_\nu(x,y)$ by our Feather combination within $AM_\nu(x,y)$ in the lower panel of this figure. Similarly, we show the variation of the A-par values as function of $R_\nu(x,y)$ in the upper panel, including the mean and sigma values (dark blue) and inter-quartile [Q10,Q90] limits (cyan) in 10 flux bins. In this representation, a perfect data combination should produce a linear correlation between the $R_\nu(x,y)$ and $I_\nu(x,y)$ fluxes (i.e. a straight line in the lower panel) with a narrow distribution independent on the reference flux (i.e. A-apar consistent with noise) in the upper panel. 

The analysis of these diagrams permits a direct evaluation of the dependence of the flux recovery with intensity in real data combination techniques
such as the one shown in Figure~\ref{fig:skymodelb_aparveflux}. Our quality assessments demonstrate how pure interferometric data fails to recover the expected reference flux distribution at all intensities (A-par $<0$ in all flux bins; see Fig.~\ref{fig:skymodelb_aparveflux}, left panel). These issues also affect regions with highest intensity and compact emission with average flux losses of ca.~25\%.
More importantly, the performance of pure interferometric data processed with CLEAN shows a strong dependence of A-par on flux, producing worse results at lower signal values. In comparison, Feather shows a clear overall improvement on the flux distribution recovery at all intensities with A-par values $\approx 0\pm 0.5$ in all our signal bins, as expected from the analysis of the total flux recovery ($A=-0.07$ or -7\% mean offset, see Fig.~\ref{fig:skymodelb_apar_all}). Despite these improvements, our analysis reveals an increase of the dispersion of A-par values at lower signal values indicating that, although better than a pure CLEANed interferometric data, the performance of Feather may be hampered in regions of low intensity emission (see a more detailed discussion in \S~\ref{sec:discussion}).

The enhanced image quality indicated by A-par both globally (Fig.~\ref{fig:skymodelb_apar_all}) and per signal bin (Fig.~\ref{fig:skymodelb_aparveflux}) corresponds to an improved flux recovery by a combination technique, in this case Feather (e.g. see Fig.~\ref{fig:skymodelb_apar_map}), compared to interferometric-only maps. An ideal data combination should in principle maximize the flux recovery at all scales. In practice, however, this may not be possible or may not be critical since, depending on the emission distribution, not all signal bins may contribute equally to the total flux budget. This is illustrated in Figure~\ref{fig:skymodelb_apar_all} (right panel) where we show the percentage of flux recovered per signal bin independently (upper sub-panel) and cumulatively (lower sub-panel) respect to the reference image. The dashed grey line in the upper sub-panel indicates the contribution of each individual signal bin to the total flux budget. \reply{We point out that for the skymodel data,} bins with intensities of $>0.2$ (in image units) contribute to most of the emission within the assessment area ($>80$~\% of the total). Recovering the flux in those bins therefore guarantees the recovery of a significant fraction of the total image sky brightness temperature. The analysis of these plots also highlights the poor performance of int-CLEAN producing negative flux values (due to PSF sidelobes plus filtering effects) in regions with extended emission showing emission values $<1$ (in image units) and leading to flux recovery values of only 25\% of the total flux.


The high sensitivity of the above comparisons demonstrates the robustness of our assessment metrics. Additional tests indicate that deviations at different flux intensities may also reveal potential issues in the data combination process (e.g. recovery of extended emission) and/or limitations of such procedures (e.g. lack of baselines overlap). Moreover, this analysis permits to investigate the performance of different data combination techniques and their response to images with different dynamic range in emission (see discussion of the data combination results in \S \ref{sec:discussion}).

\subsection{Power Spectra (PS): spatial scale sensitivity}
\label{sec:powerspec}

\begin{figure}[ht]
\centering
\includegraphics[width=0.7\textwidth]{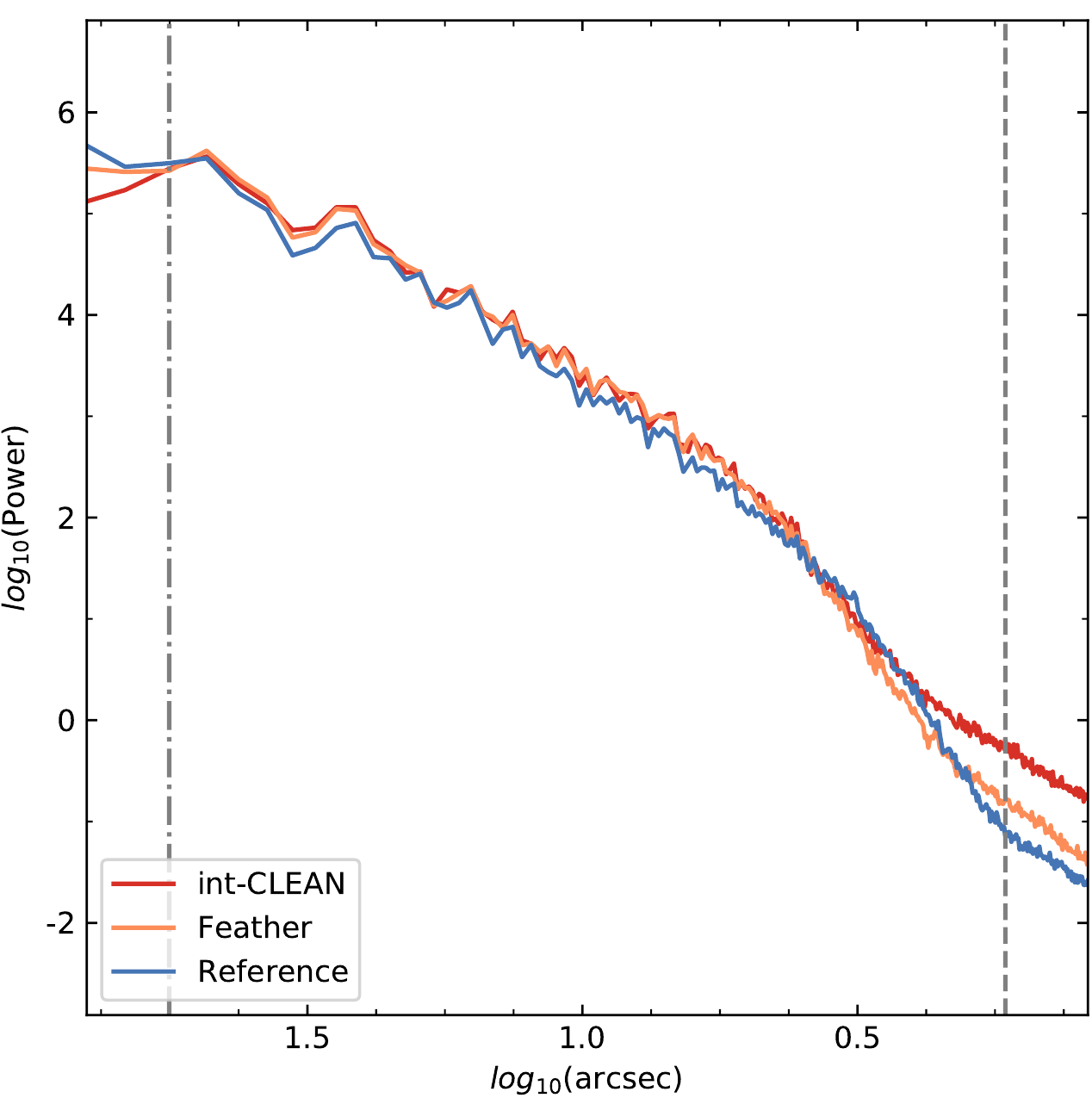}
\caption{\label{fig:skymodelb_ps} Spatial Power Spectra (SPS) obtained from the pure interferometric data (int-CLEAN, red) and the Feather-combined interferomentric and SD data (orange) compared with the original power spectrum of our skymodel simulation (dark blue). Note the logarithmic scales and the several orders of magnitude of flux present in the input model image. Vertical lines show \reply{the final image resolution} (dashed line, right) and largest angular scale effectively sampled by the SPS (1/3 of  the map size; dot-dashed line, left), beyond which this simulated instrument is not sensitive.}
\end{figure}

The primary goal of each data combination method is the successful recovery of emission at the angular scales filtered out by the interferometer. Thus, it is crucial to describe how the recovered emission is distributed as a function of angular scale. One such tool is the spatial power spectrum (SPS) defined as
\begin{equation}\label{eq:SPS}
P\left(k\right) = \mathscr{F}\left[I_\nu\left(x,y\right)\right]\mathscr{F}^*\left[I_\nu\left(x,y\right)\right], 
\end{equation}
where $k$ is the wavenumber ($k = \frac{1}{\theta}$ where $\theta$ is the angle on the sky in rad) and $\mathscr{F}$ is the Fourier transform of the input image, $I_\nu\left(x,y\right)$, under study. This is a two-point correlation function that measures how power (i.e., structure) is distributed across spatial scales. In practice the distribution of power is measured by computing the 2D Fourier transforms of the integrated intensity images and measuring the median in progressively larger annuli. We chose the median to mitigate the bias introduced by ringing artifacts of the \textit{sinc} function along the axis of the 2D FT when emission extends toward the edge of the image. While phase information is lost due to the multiplication by the complex conjugate, comparing the SPS profiles derived from the images produced by the various combination methods to the input reference (or total power) demonstrates how successfully the emission is recovered across the full range of angular scales. The ideal image would recover the flux of the input model image on all scales.

As an example, we present in Fig. \ref{fig:skymodelb_ps} the SPS profiles measured for the data combination of our skymodel using Feather in comparison to the interferometer-only data processed with CLEAN. Also shown is the profile of the original input image of the simulation (Reference). These three profiles agree remarkably well at larger angular scales (left side of the plot). Feather and reference continue to agree all the way down to the smaller angular scales accessible to the instrument. But the int-CLEANed interferometer-only data start to diverge from the reference as the scale size decreases. 
\reply{This result can be understood by considering the way the power is measured: }
The insufficient recovery of the larger scales effectively boosts \reply{the intensity contrast of} the already existing small-scale structure in the interferometer-only image \reply{leading to \textit{higher} power (i.e. larger differences) at small angular scales compared to the reference image}.
Conversely (not shown in this example), when the SD data is given too much weight in the combination, the power \textit{decreases} at small scales because the low resolution SD data effectively washes out the observed small-scale structure. 
The variation in power at small scales highlights subtle deficiencies that may not be immediately and clearly visible in the images. Scales below the beam size (right of the dashed line in the figure) are dominated by correlated noise in the pixels or convolution residuals and should generally be ignored.

\begin{figure}[ht]
\centering
\includegraphics[width=0.7\textwidth]{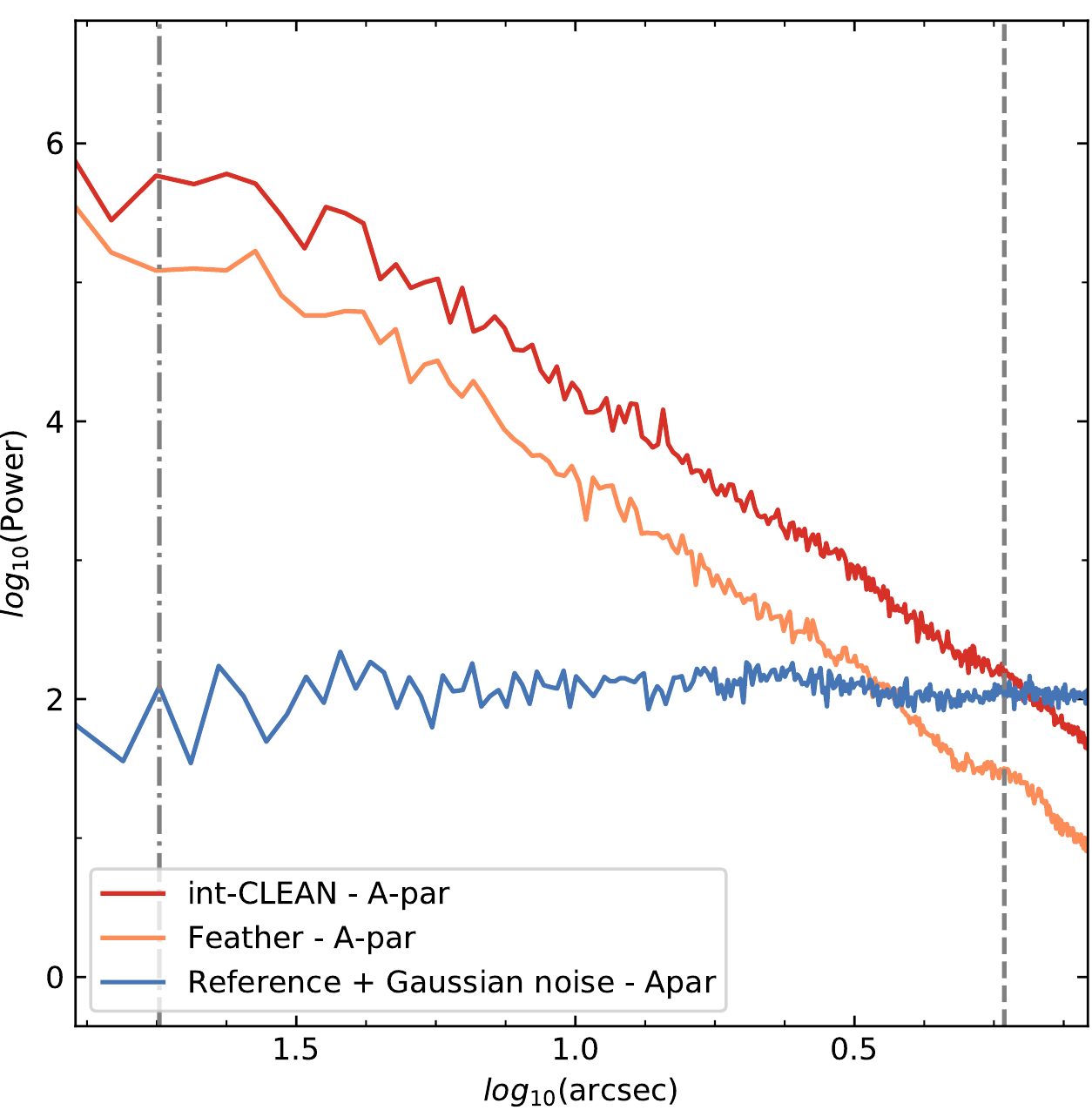}
\caption{\label{fig:skymodelb_Aparps} A-par Spatial Power Spectra (A-par SPS) for the same int-CLEAN and Feather data reductions shown in Fig.~\ref{fig:skymodelb_ps}. For comparison we add the expected (flat) A-par power spectrum of our reference image with Gaussian noise added. 
}
\end{figure}

The above SPS analysis can also be applied to the case of our A-par maps (e.g. Fig.~\ref{fig:skymodelb_apar_map}, lower left panel). The expected A-par power spectrum provides information on the variance of the relative errors, $\sigma$(A-par), as a function of scales (e.g. high A-par power denotes large systematic errors at a given scale). In addition to the minimization of the (global) A-par mean and dispersion values (e.g. in Fig.~\ref{fig:skymodelb_apar_all}), the goal of any data combination method is therefore to minimize the  A-par SPS, that is, all systematic errors as a function of scale. We show the results of the A-par SPS for our int-CLEAN and Feather reductions in Fig.~\ref{fig:skymodelb_Aparps}. Our results illustrate the reduction of the A-par variance at all scales once a Feather combination is applied to the original int-CLEAN reduction. The interpretation of these results in comparison with other combination methods and the addition of Gaussian noise to our reference skymodel is discussed in \S~\ref{sec:methods_SPS}.

\subsection{Using single-dish observations as reference image} \label{sec:comp-w-sd}

Unlike the case of our synthetic Skymodel as presented in our previous assessments, the quality assessment of real scientific datasets is usually hampered by the lack of information for reference at high spatial resolution. Instead, when considering total emission on all scales, interferometric observations can only be cross-checked with the flux measurements obtained at SD resolution. 
Such comparison requires the original interferometric image at its native (high-)resolution $I_\nu(x,y)|_0$ to be convolved into a final (low-)resolution image $I_\nu(x,y)$ similar to the SD data $R_\nu(x,y)$.

\begin{figure}[t]
\includegraphics[width=0.5\textwidth]{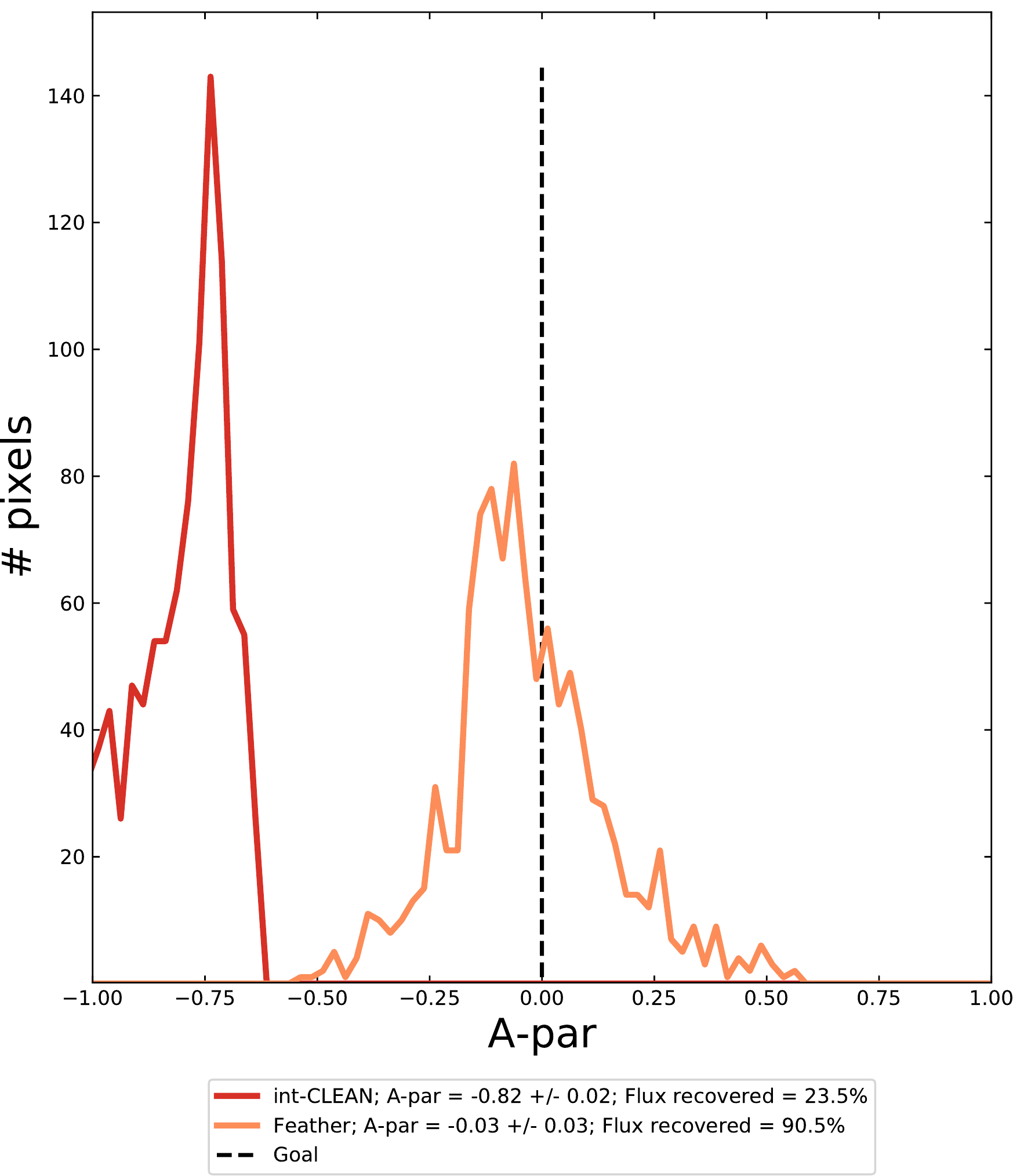}
\includegraphics[width=0.5\textwidth]{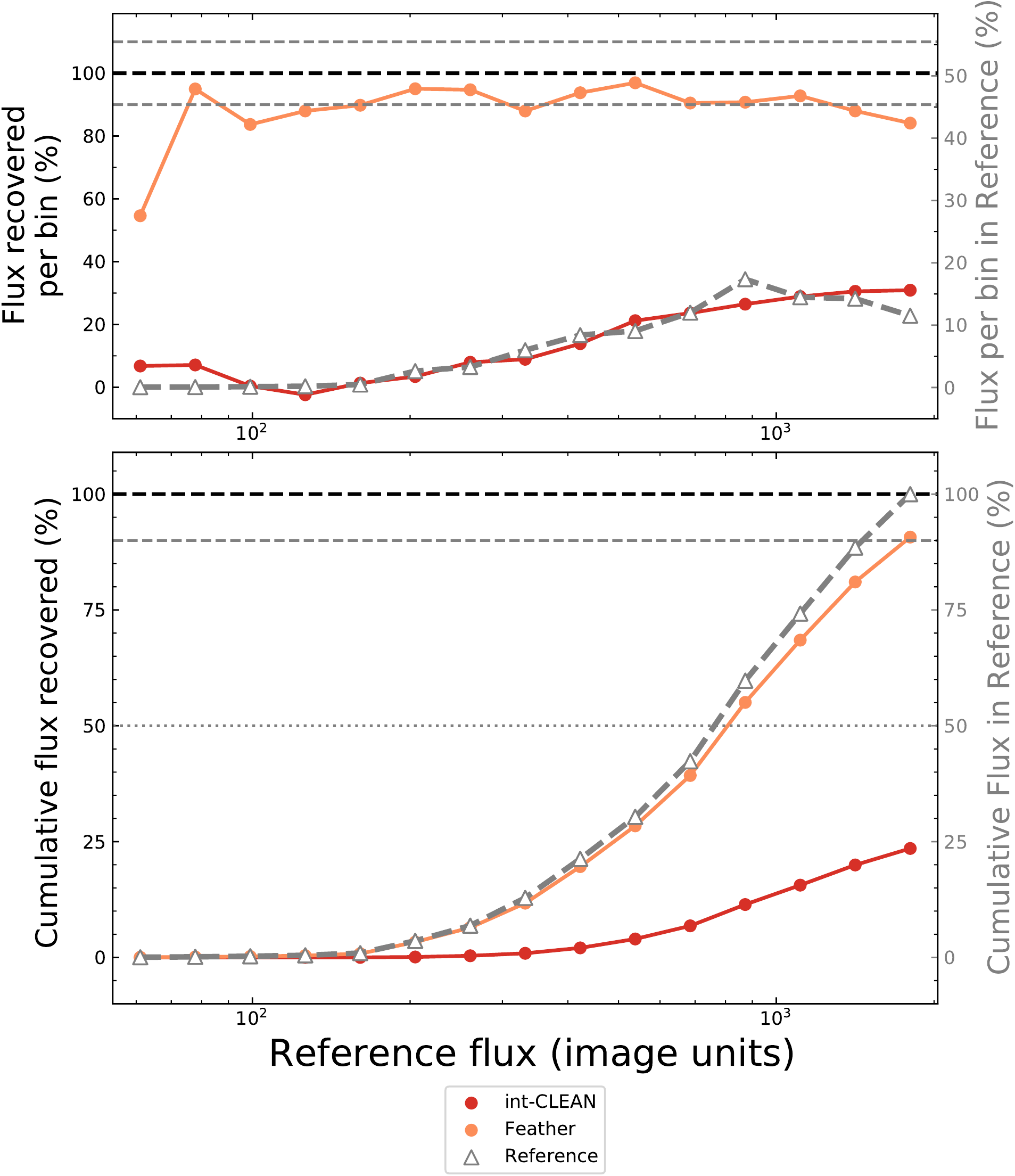}
\caption{Same as Figure~\ref{fig:skymodelb_apar_all} but using the SD image as reference rather than the simulation input image as it would be done when working with real observational data.
\label{fig:sd_apar_all}}
\end{figure}

We quantify the effects of this low-resolution comparison in our assessment metrics in Figure~\ref{fig:sd_apar_all} showing the results of int-CLEAN (red) and Feather (blue) for our skymodel simulation, this time convolved into the SD resolution (57\arcsec), for the case of the A-par statistics (left panel) and the recovered flux-per-signal-bin (right panel). Figure~\ref{fig:sd_apar_all} can be directly compared with Fig.~\ref{fig:skymodelb_apar_all} carried out at much higher resolution (1.7\arcsec). Several results can be taken from this comparison. The number of pixels for the quality assessment is largely reduced due to the use of a much larger beamsize as seen in the overall pixel statistics. Changes in the beamsize also alter the local A-par estimates due to the averaging of regions with/without emission within the much smoother SD beam.
Nonetheless, our assessment metric preserves the relative differences between methods. Both the A-par statistics (left panel) and the recovered flux (right panel) indicate the improved performance of data combination techniques such as Feather in comparison with interferometric-only CLEAN deconvolution.  

Several caveats should be considered during the quality assessment of images at the SD resolution. First, and by construction, the SD observations only provide information of the total flux per SD beam but the emission distribution at smaller scales remains unconstrained.  
Second, large convolutions (such as the one shown in Fig.~\ref{fig:sd_apar_all}) can lead to averaging effects of multiple emission features at interferometric resolutions (e.g. compact sources and negative sidelobes) unresolved at the SD beamsize.
Third, the limited amount of resolution elements in the SD images restricts the use of some size-dependent assessments such as SPS.
Treated with caution at low-resolutions, however, our assessment metrics provide a powerful toolkit to assess the quality of both continuum and spectral interferometric observations. 

\subsection{A-par spectrograms: analysis of spectral cubes}\label{sec:spectrograms}

\begin{figure}[t]
\centering
\includegraphics[width=1.\textwidth]{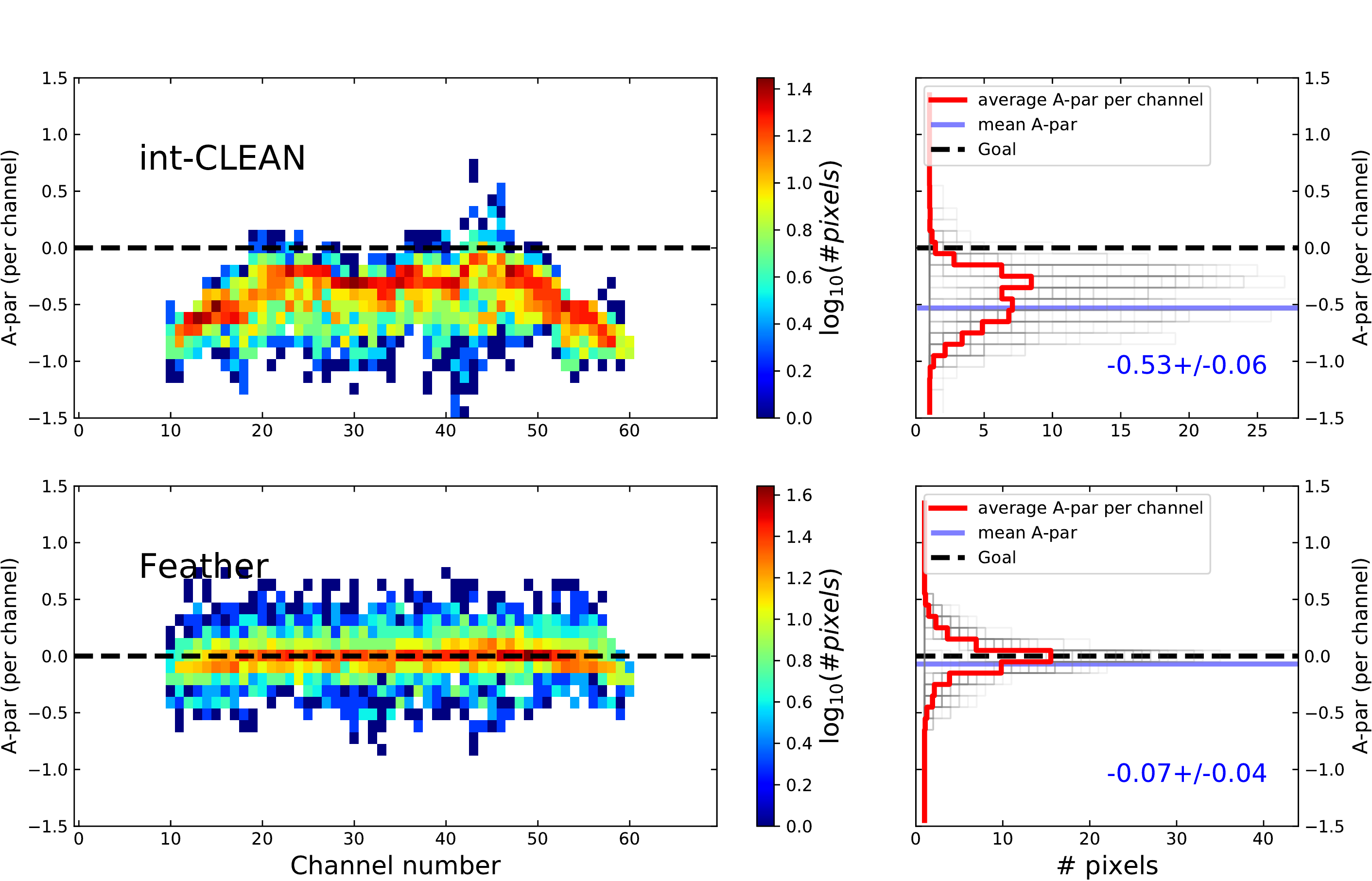}
\caption{A-par ``spectrogram'' for our M100 data obtained from the interferometric data alone with int-CLEAN (top) and from the Feather combined interferometric data and SD image cube (bottom) for all voxels within our assessment mask. The detected emission in each spectral channel of this CO line observation represents emission from gas at a specific velocity range with its own spatial distribution.    \label{fig:m100_apar}}
\end{figure}

In order to extend the previous assessments, which were of a continuum (single channel) image, to a data cube, we must visualize the variation in the spatial distribution as a function of frequency (or velocity) channel.  This is important because the astronomical objects are likely to have distinct morphology for material at different velocities.
Hence, any combination method, or perhaps no combination, might be sufficient for the channels with more compact emission, whereas the channels with more extended emission depend much more on the combination for an acceptable image quality. 

We extend the A-par analysis from the previous continuum case (e.g. Figure \ref{fig:skymodelb_ps}) to that of a cube in the form of an ``A-par spectrogram'', shown in Figure \ref{fig:m100_apar}. This spectrogram shows the distribution of A-par values for all the channels in the target cube that satisfy our assessment mask $AM_\nu(x,y)$ criteria (left panel). For a given velocity channel (x axis), the spectrogram shows the number of voxels within this frequency plane (color scale) with a given A-par value (y axis). Its analysis provides a visual comparison of the emission recovered in multiple channels and allows the evaluation of local (per channel) and global (mean distributions) deviations as well as issues (e.g. outliers  and/or problematic channels) in the data.
Similar to the analysis of the continuum images, the goal of any data combination method is to minimize the value of this ``A-par spectrogram'' for all channels within a cube.
We complement these results with additional histograms (right panel) for A-par showing all individual channel values (grey histograms) as well as the mean distribution (red histogram). 

These spectrograms turn out to be a powerful and intuitive tool when comparing multiple data combination techniques in large spectral cubes. In Fig.~\ref{fig:m100_apar} (top panel) we illustrate again the limitation of the interferometer-only emission recovery (A-par$<$~0) at different velocities/channels. Moreover, our spectrograms illustrate how the flux losses also depend \replyb{on} velocity, \reply{since the brightness distribution also varies with velocity (see \ref{sec:lineprofile})}. Depending on the selected channel, A-par varies between $\sim$~-0.1 (10\% flux losses) and $>$~-0.5 ($>$~50\% flux losses). In contrast, data combination with Feather (lower panel) produces significantly better results at all velocities/channels with typical values of $|$A-par$|\sim$~0.2 (or $\pm$~20\% error) for most channels (between $[10,60]$). The improvement is also seen in the mean A-par vaues in the corresponding histograms. The broad and shifted distribution in the int-CLEAN reduction with mean 50\% flux losses (A-par$=-0.51\pm0.07$) is significantly improved by Feather showing a narrower and more centered distribution showing differences of less than 10\% respect to the SD flux per channel (A-par$=-0.07\pm0.04$).

A variation of A-par with channel directly translates into a deformation of spectral line features as will be discussed in \S \ref{sec:lineprofile}. That is the case for our int-CLEAN reduction as seen in Fig.~\ref{fig:m100_apar} (top panel). In other words, data combination is also relevant for spectroscopy.

\section{Combination Method Comparison and Discussion} \label{sec:discussion}

\subsection{The indispensable short-spacing information}\label{sec:short_spacing}

\begin{figure}[t]
\includegraphics[width=1.\textwidth]{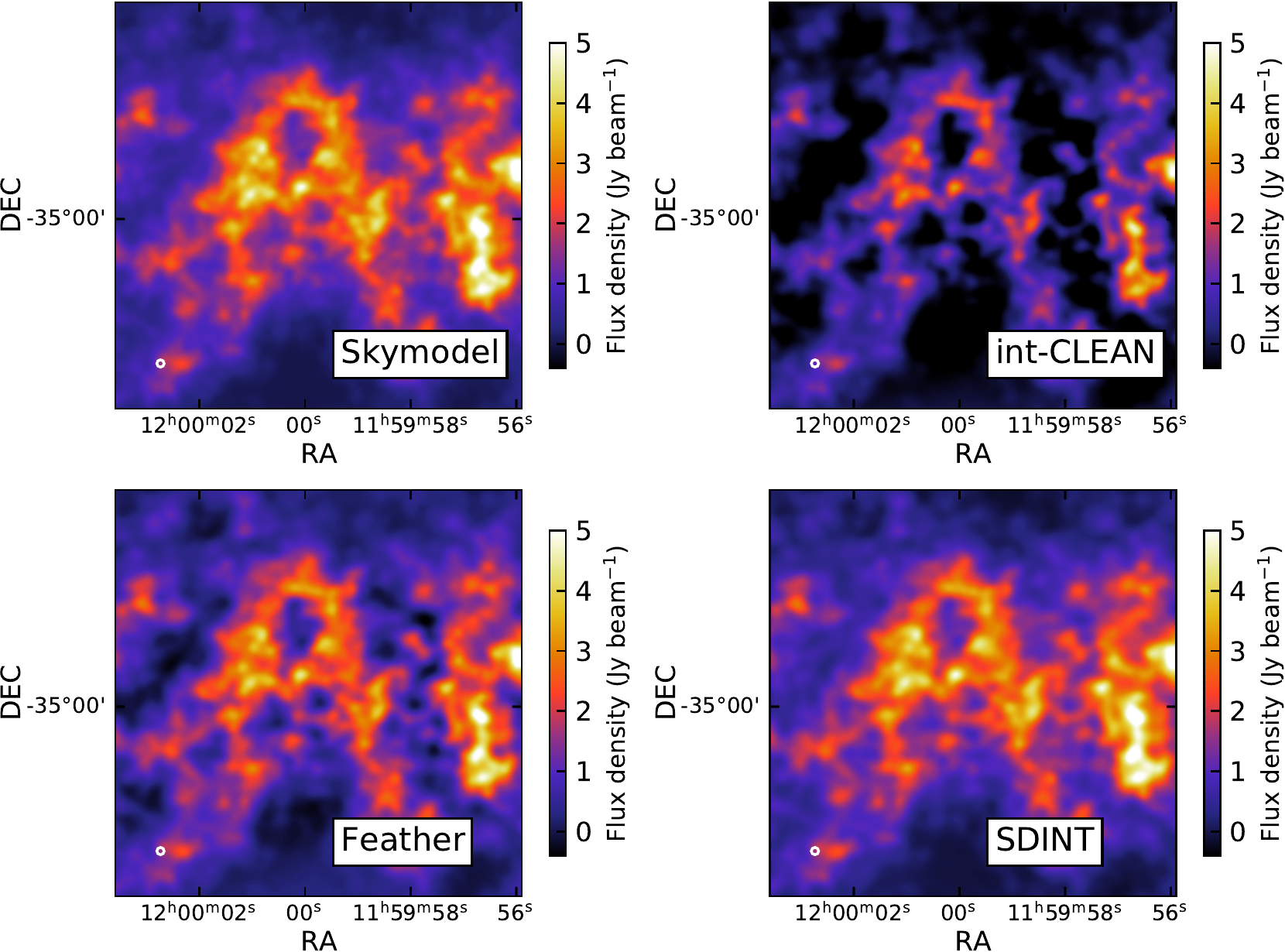}
\caption{Comparison of the true sky brightness distribution with the recovered emission of interferometry-only and two different data combination methods in the central region of our Skymodel: (top left) Skymodel image (i.e. reference), (top right) int-CLEAN (i.e. only interferometric data), (bottom left) Feather, and (bottom right) SDINT, all convolved to a common resolution of 1.7\arcsec\ (see beamsize in the lower left corner) and shown with the same intensity scale. \label{fig:clean_vs_DC}}
\end{figure}

The results presented in \S~\ref{sec:assess} demonstrate the large impact of the short-spacing information on the analysis of interferometric observations. All assessment metrics show significant improvements of the flux recovered at different spatial scales and input signals using a standard data combination technique such as Feather when compared to pure interferometric-only CLEAN (int-CLEAN) reductions (see Figs.~\ref{fig:skymodelb_apar_map}-\ref{fig:m100_apar}). 
Here, we explore the relative performance of additional data combination techniques.

\begin{figure}[t]
\centering
\includegraphics[width=0.9\textwidth]{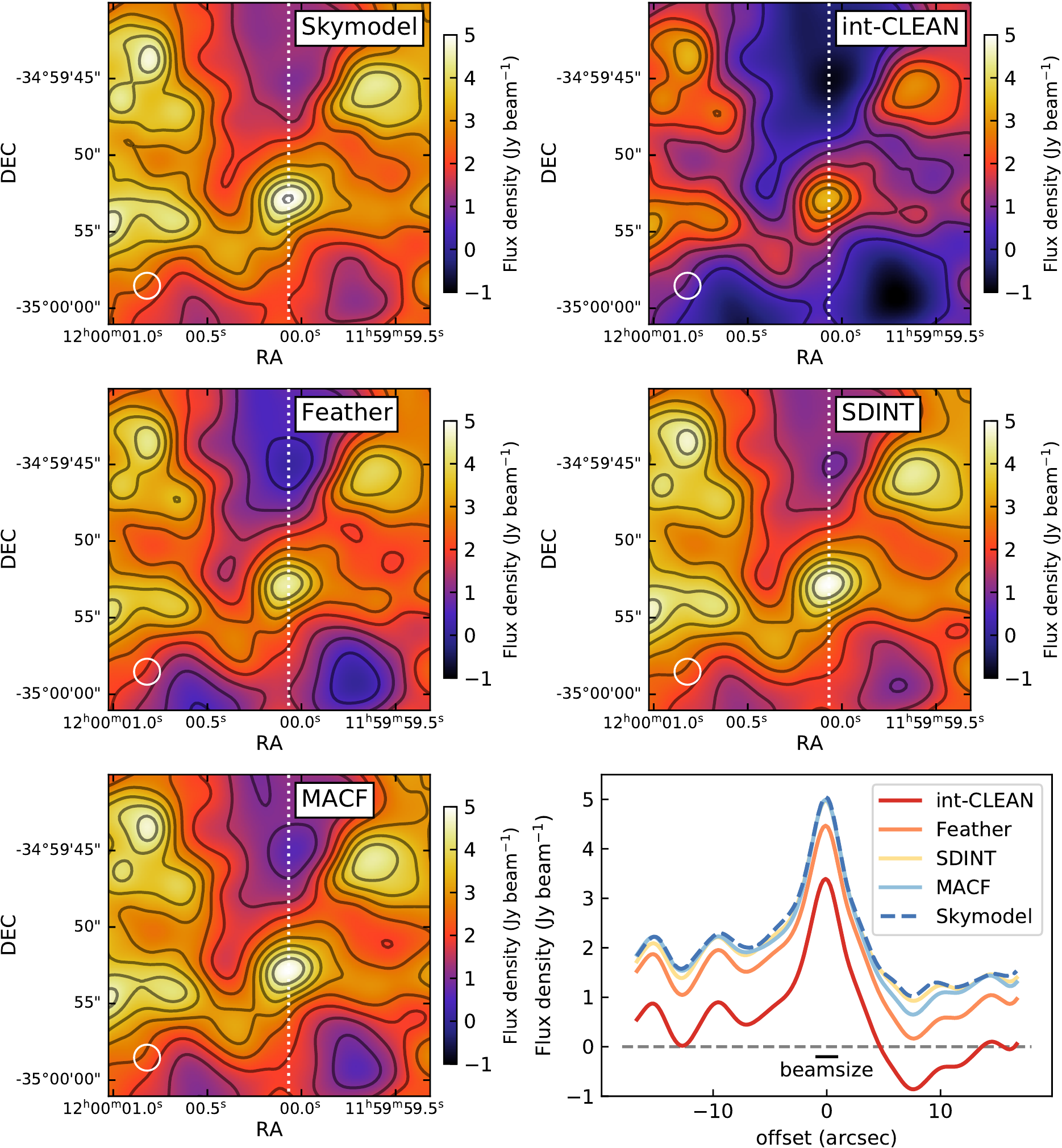}
\caption{Compact emission recovered by different data combination methods at the central region of skymodel image all convolved into a final resolution of 2~\arcsec (see bottom left corner of each image) and plotted within the same intensity range and with the same contours every 0.5~Jy~beam$^{-1}$. From left to right and from top to bottom: skymodel, Int-CLEAN, Feather, SDINT, and MACF results.  The lower right panel shows the intensity profile indicated with a dotted line in our maps obtained by these methods. \label{fig:compact}}
\end{figure}

We highlight the enhanced image quality of different data combination techniques in Figure~\ref{fig:clean_vs_DC} by comparing our input Skymodel emission (reference; top left panel) with the results obtained by int-CLEAN (top right panel), Feather (bottom left panel), and SDINT (bottom right panel) in the central region of the Skymodel image (see also Fig.~\ref{fig:skymodelmaps}; FSSC and MACF maps are not shown as they produce visually similar results to Feather and SDINT, respectively). This region is selected because it incorporates both compact and extended emission.
A direct inspection of these images reveals how pure int-CLEAN deconvolutions (top right panel) miss large fractions of the true sky emission at all scales and flux values due to the incomplete sampling of the \emph{(u,v)}-space in interferometric-alone observations when extended sky emission is present.  The resulting interferometric filtering becomes non-intuitive, showing highly non-linear effects in the case of complex sources and leading to a flux recovery of only $\sim$~10-20\% of the true sky emission (see Fig.~\ref{fig:skymodelb_apar_all}, red lines). These effects can be sometimes identified in our images by the appearance of regions with negative emission (negative sidelobes; dark areas in the int-CLEAN image)\footnote{We note here that the absence of negative emission troughs in the image does not guarantee the absence of filtering, as the effects of filtering can be compensated by the presence of additional positive emission depending on the source structure.}.  

Interferometric-alone observations (int-CLEAN) are expected to efficiently recover the total flux of unresolved, isolated sources with sizes comparable to the interferometer beamsize. However, \reply{this situation is not the same in the presence of smooth, extended emission.}  In the latter cases, it is sometimes assumed that the lack of short-spacing information only affects the recovered emission beyond the Maximum Recoverable Scale (MRS)\footnote{See  \cite{remijan2019} for a definition. The MRS depends on the details of the interferometric array setup.  For the ALMA 12m nominal configurations with 43 antennas, the MRS is approximately a factor 10 larger than the achieved angular resolution.} sampled by the interferometer. But our results clearly demonstrate how the interferometric filtering has a significant impact at all scales including those compact regions with sizes comparable to the beamsize (i.e. angular scale $\ll$~MRS). 

The addition of the short-spacing (SD) information dramatically improves the image quality. Standard techniques such as Feather (Fig.~\ref{fig:clean_vs_DC}, lower left panel) recover up to $\sim$~90\%  of the input Skymodel emission (see Fig.~\ref{fig:skymodelb_apar_all}, blue lines) and remove most of the negative emission patches in the field of interest. 

The addition of the short-spacing information at low-resolutions also has a direct impact on the peak flux in many of the compact sources in this region, potentially affecting column density and mass estimates in subsequent scientific analysis.  We demonstrate these issues in Fig.~\ref{fig:compact} by comparing the emission recovered by int-CLEAN, Feather, SDINT, and MACF at the central region of our skymodel showing multiple compact emission features with sizes comparable to the beamsize. For illustrative purposes, we compare the emission profile of the brightest region within this field (see lower right panel). Although compact, the peak flux of this source in the interferometric-alone int-CLEAN observations is reduced by a factor of $\sim$~40\% compared to the true emission in our skymodel. Flux losses are also seen in other nearby sources together with negative sidelobes in regions with fainter emission (see also upper right panel). Still, the effects on our maps depend on the location of the source and the structure of their surroundings, making impossible a direct correction without the zero-spacing information.
The use of data combination techniques significantly improve the flux recovery in these compact regions with, in our particular case, SDINT and MACF producing the best results. These reported differences could have a significant impact on derived properties such as mass and column densities of compact sources in complex environments. These results underline the absolute need for short-spacing information and data combination in the analysis of particular high-dynamic range observations provided by state-of-the-art interferometers such as ALMA.

\subsection{Effects of the short-spacing information in line profiles}\label{sec:lineprofile}

\begin{figure}[t]
\centering
\includegraphics[width=0.9\textwidth]{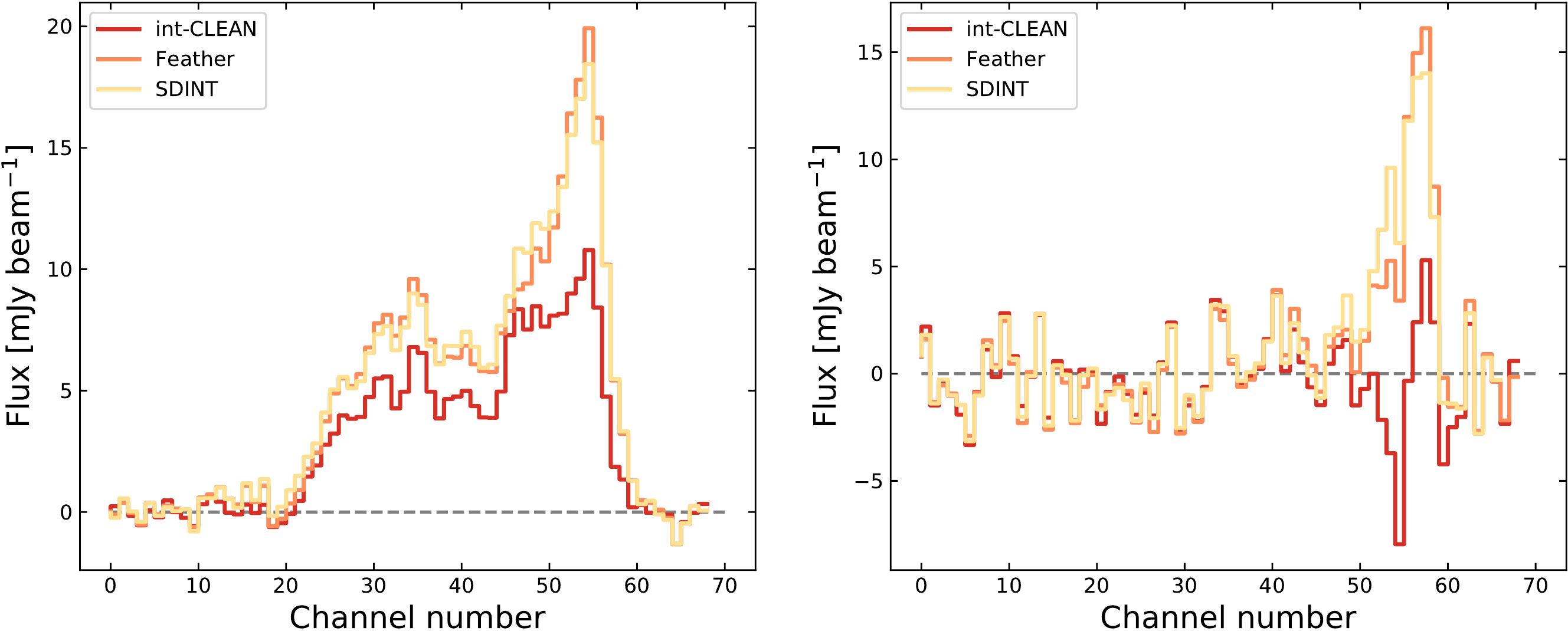}
\caption{Examples of the data combination on lines profiles obtained from our analysis of two representative regions of our M100 dataset: (Left panel) Line profile in one region shows effects on the line peaks and relative intensities; (Right panel) Line profile in another, lower-flux region shows effects on the line shape. \label{fig:M100spectra}}
\end{figure}

So far the discussion focused on the analysis of the integrated emission maps. As already introduced in \S~\ref{sec:spectrograms}, \reply{the extent of the effects of lack of short-spacing information will depend on the target structure in each frequency channel.} We illustrate these effects in Fig.~\ref{fig:M100spectra} in two representative regions in our M100 dataset. 
As shown in Fig.~\ref{fig:M100spectra} (left panel), the interferometric filtering produces changes of up to 50\% on the flux per channel in this spectrum. The true spatial structure of the emission and thus the fraction of missing observed flux is however not necessarily homogeneous but can vary from channel to channel. According to Fig.~\ref{fig:m100_apar} (upper panel), these losses can be as high as 90\% in some channels.  This can alter line ratios by several ten percent, an effect already reported in previous studies \citep[e.g.,][]{Pety2013}.

Figure~\ref{fig:M100spectra} (right panel) shows how frequency-dependent, spatial filtering effects can also modify the line Full Width at Half Maximum (FWHM) and line centroid of the observed gas components. As illustrated in the int-CLEAN spectrum,
negative sidelobes can selectively alter different channels (see negative feature in the spectrum) both reducing the line FWHM and shifting the line peak in comparison with the Feather and SDINT reductions.
Data combination is therefore essential to recover not only the total flux in spectral cubes but also to preserve the kinematic information in them. For \replyb{wider} frequency coverage, one should also ensure that the frequency dependence of \emph{(u,v)} spacing and PB are taken into account.

\subsection{Comparison between data combination methods}\label{sec:method_comparison}

\begin{figure}[ht]
\includegraphics[width=0.5\textwidth]{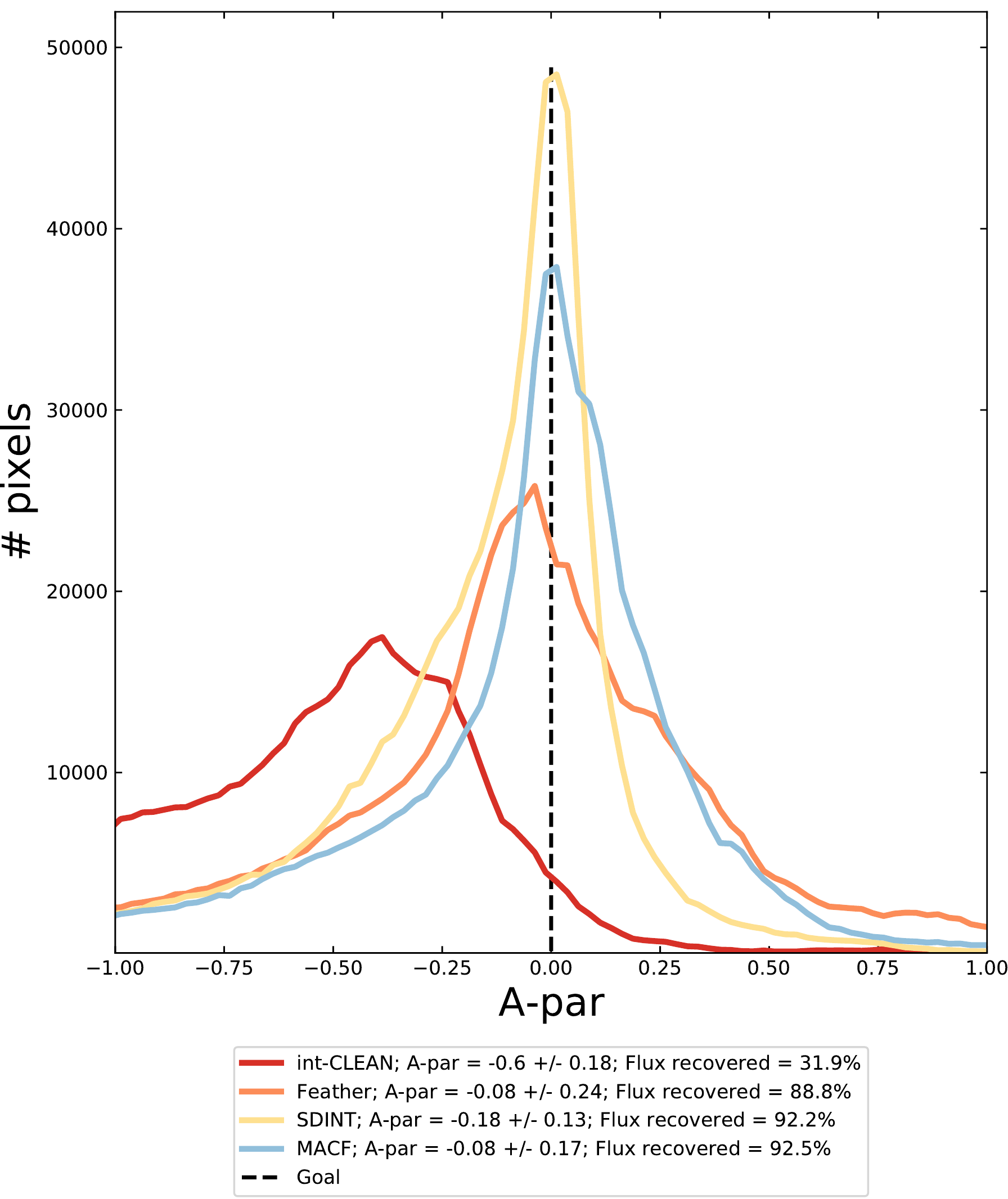}
\includegraphics[width=0.5\textwidth]{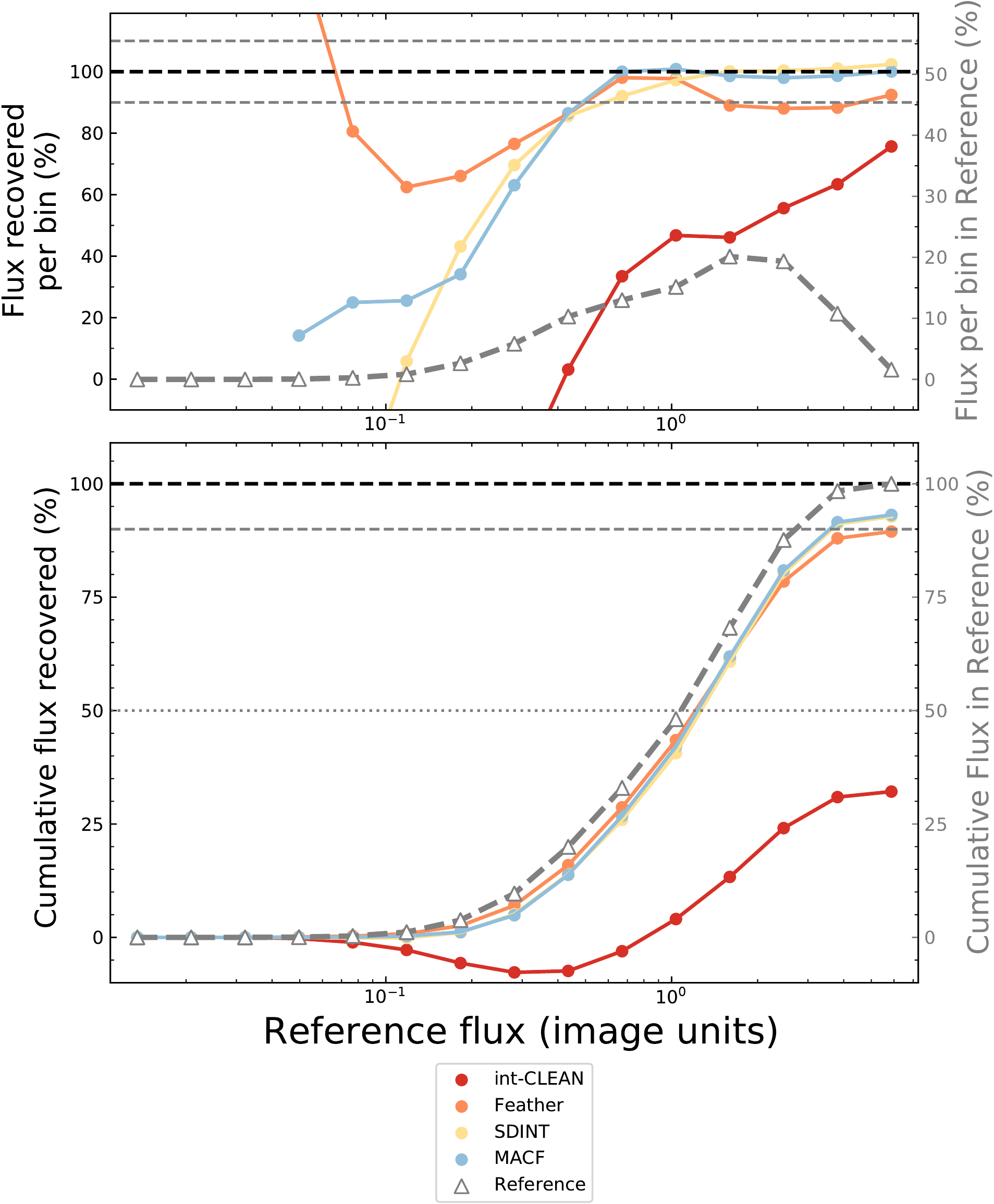}
\caption{Comparison of the data combination methods for the case of our Skymodel: int-CLEAN (red), Feather (orange), SDINT (yellow), and MACF (light blue). (Left panel) A-par statistics. (Right panel) Flux recovered by flux bin \reply{compared to our skymodel as reference (grey triangles)}. See also Fig.~\ref{fig:skymodelb_apar_all}, as gray \reply{horizontal} lines are the same.\label{fig:allmethods}}
\end{figure}

Inspecting Fig.~\ref{fig:clean_vs_DC} in detail suggests that more elaborate methods such as SDINT (lower right panel) may be able to reproduce the original sky brightness distribution (upper left panel) even better than Feather. 
We quantitatively investigate the performance of the different data combination techniques introduced in \S~\ref{sec:methods}, all of which are available in the script we provide. In Fig.~\ref{fig:allmethods} (left panel) we show the overall A-par values as well as the total recovered flux in the images (see also figure caption) obtained by the int-CLEAN, Feather, SDINT, and MACF methods, as applied to the Skymodel. The FSSC method is not shown but produced results nearly identical to Feather in this test. Feather recovers $\sim$~89\% of the total flux with a mean accuracy of A-par~$=-0.08$ and an accuracy dispersion of $\sigma($A-par$)=0.24$. The total flux recovery is improved
by the SDINT and MACF methods with total flux values above $\sim$~92\% with respect to the true sky emission. While both are better than Feather, we notice quantifiable differences between the SDINT and MACF methods. While SDINT shows a smaller dispersion of the A-par values ($\sigma($A-par$)=0.13$, seen as a narrower distribution in this histogram), MACF shows a better mean A-par value (A-par~=~-0.08, leading to a more centered distribution). These differences are likely connected to the slightly different responses of these methods to the flux intensity (see Figure~\ref{fig:allmethods}, right panel). Compared to Feather (and FSSC), both SDINT and MACF maximize the flux recovery in those high intensity flux bins contributing to most of the total sky emission.

\subsection{Flux and A-par power spectra}\label{sec:methods_SPS}

\begin{figure}[ht]
\centering
\includegraphics[width=0.8\textwidth]{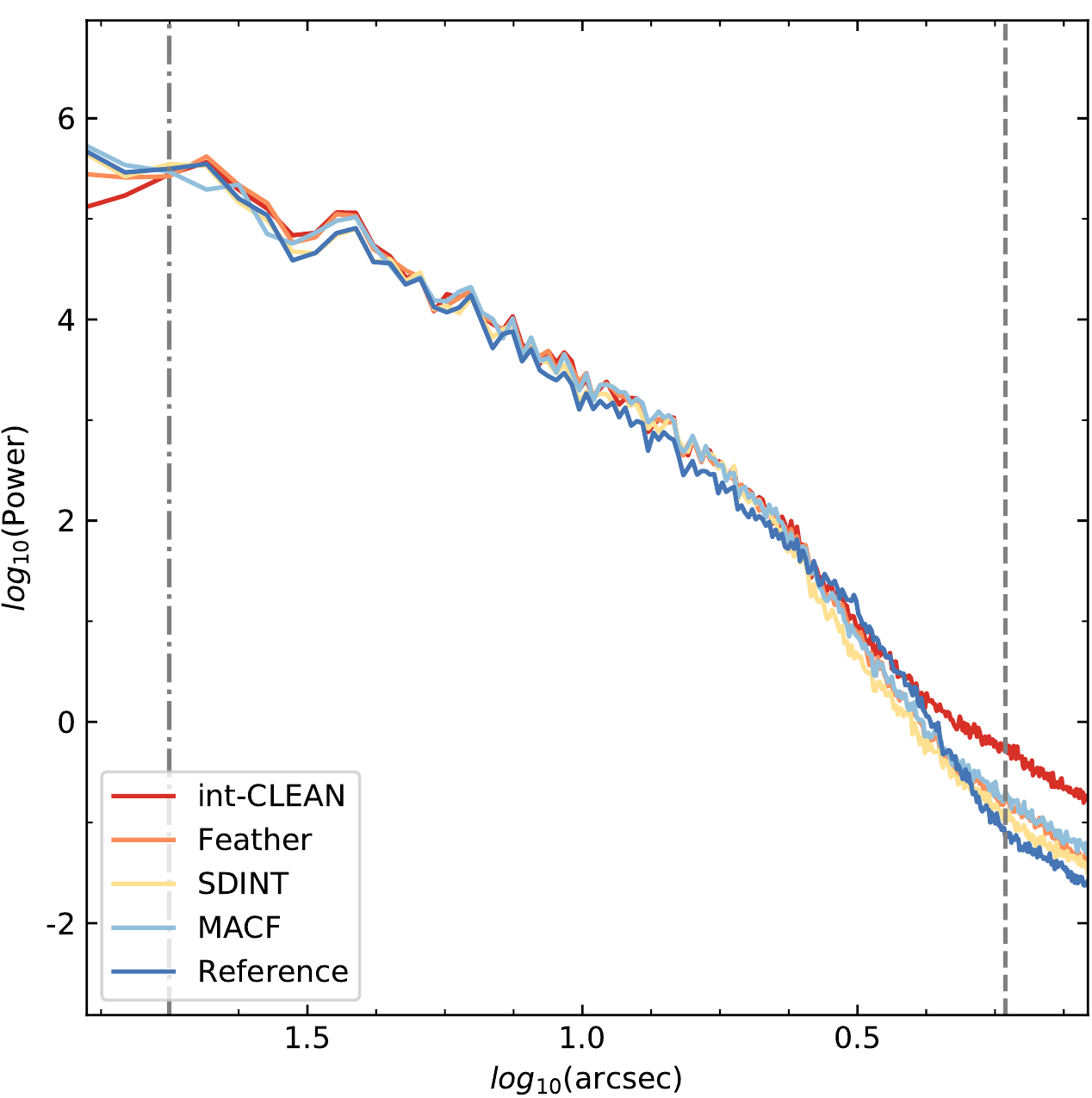}
\caption{\label{fig:allmethods_PS}Power spectrum of image intensity as introduced in \S~\ref{sec:powerspec} and Fig.~\ref{fig:skymodelb_ps}, here also showing the results for SDINT (yellow) and MACF (light blue) \reply{with our skymodel as reference (dark blue)}. As in Fig.~\ref{fig:skymodelb_ps}, vertical lines show the range above and below which this simulated instrument is not sensitive.}
\end{figure}

\begin{figure}[ht]
\centering
\includegraphics[width=0.8\textwidth]{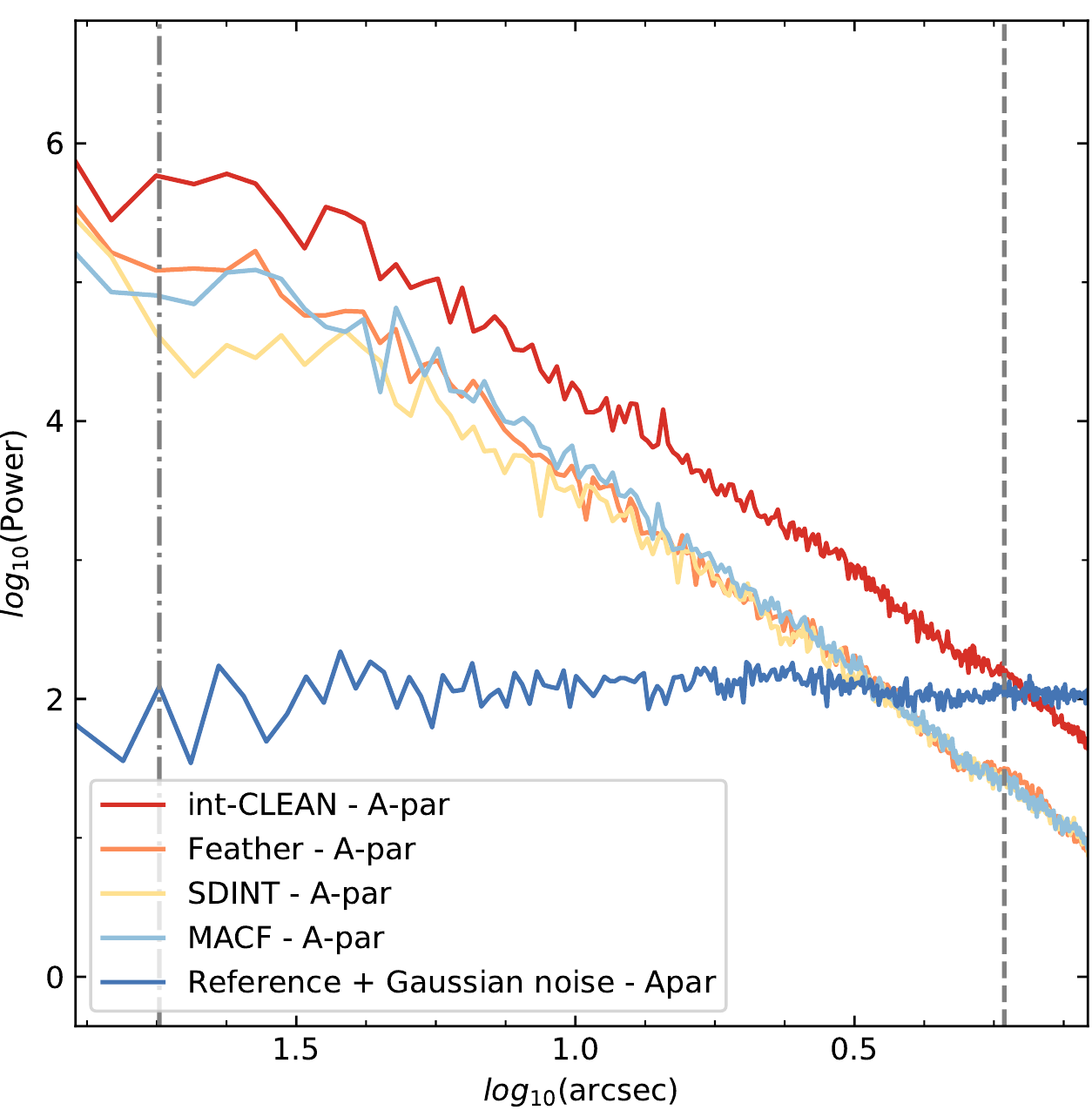}
\caption{\label{fig:allmethods_AparPS}Power spectrum of the A-par map as introduced in \S~\ref{sec:powerspec} and Fig.~\ref{fig:skymodelb_Aparps}, here also showing the results for SDINT (yellow) and MACF (light blue).}
\end{figure}

Continuing the discussion in \S~\ref{sec:powerspec} we show in Figs.~\ref{fig:allmethods_PS} and \ref{fig:allmethods_AparPS} the intensity power spectrum and the A-par power spectrum for the results on our simulated Skymodel observation obtained by interferometry only and the different combination methods.

The intensity power spectrum (Fig.~\ref{fig:allmethods_PS}) of all combination methods traces quite closely that of the reference at first glance but with small variations. Between the largest scales sampled by our SPS (left vertical dot-dashed line) and log(angular\ scale)$\sim$0.7, all results are on or above the reference with SDINT showing the closest power. Between log(angular\ scale)$\sim$0.7 and $\sim$~0.4, the results are constantly below the reference. Here, MACF is closer to the reference. Shortly before reaching the smallest accessible angular scales (right vertical dashed line) the results rise again above the reference. And in this range SDINT performs best. Although tracing scales rather than only flux recovery, the results of these SPS are consistent with the analysis of Fig.~\ref{fig:allmethods} (right panel).

The A-par map power spectra shown in Fig.~\ref{fig:allmethods_AparPS} speak a clearer language because in this type of plot, the deviations from the reference occur in the angular scale range where the instrumental sensitivity is actually missing. All results with or without data combination, and no matter which method is used, are far above the reference and only come close to it at the smallest angular scales (right side of the plot). This is because the reference compared to itself of course has an A-par of zero and the non-zero A-par values in the plot are only caused by the added (Gaussian) noise which has no spatial structure. The deviations of the reconstructed images from the reference, however, do have a spatial structure, and A-par is extremely sensitive to this (see also Fig.~\ref{fig:skymodelb_apar_map}, lower left panel). In other words, the slopes seen in these A-par SPS compared to a pure white noise (dark blue line in the plot) demonstrate how the deviations of any data combination method compared to the original reference image are not Gaussian but depend on scale and intensity (see distribution of the A-par variations in Fig.~\ref{fig:skymodelb_apar_map}, lower left panel).
With this sensitivity, we can see that SDINT is performing increasingly better (lower, flatter) than the other combination methods as the angular scale increases (towards the left). 
This corresponds to the narrowness of the A-par distribution seen in the previous section.

Our experiments demonstrate how interferometers can significantly alter the original power spectrum of a science target such that in the Skymodel. The incomplete sampling of the \texttt{(u,v)}-plane information introduces non-linear effects at different scales (see A-par SPS in Fig.~\ref{fig:allmethods_PS}) modifying both the absolute value and slope of the recovered SPS that persist even after data combination (see Fig.~\ref{fig:allmethods_AparPS}). Similar results are found in other structure analysis techniques such as Probability Distribution Functions \citep[PDF; e.g. see][]{Ossenkopf2016}. The physical interpretation of the recovered SPS in interferometric observations (e.g. absolute values and changes in the power spectrum slope) should be treated with extreme caution. 

\subsection{Single-dish flux recovery}

\begin{figure}[ht]
\centering
\includegraphics[width=1.\textwidth]{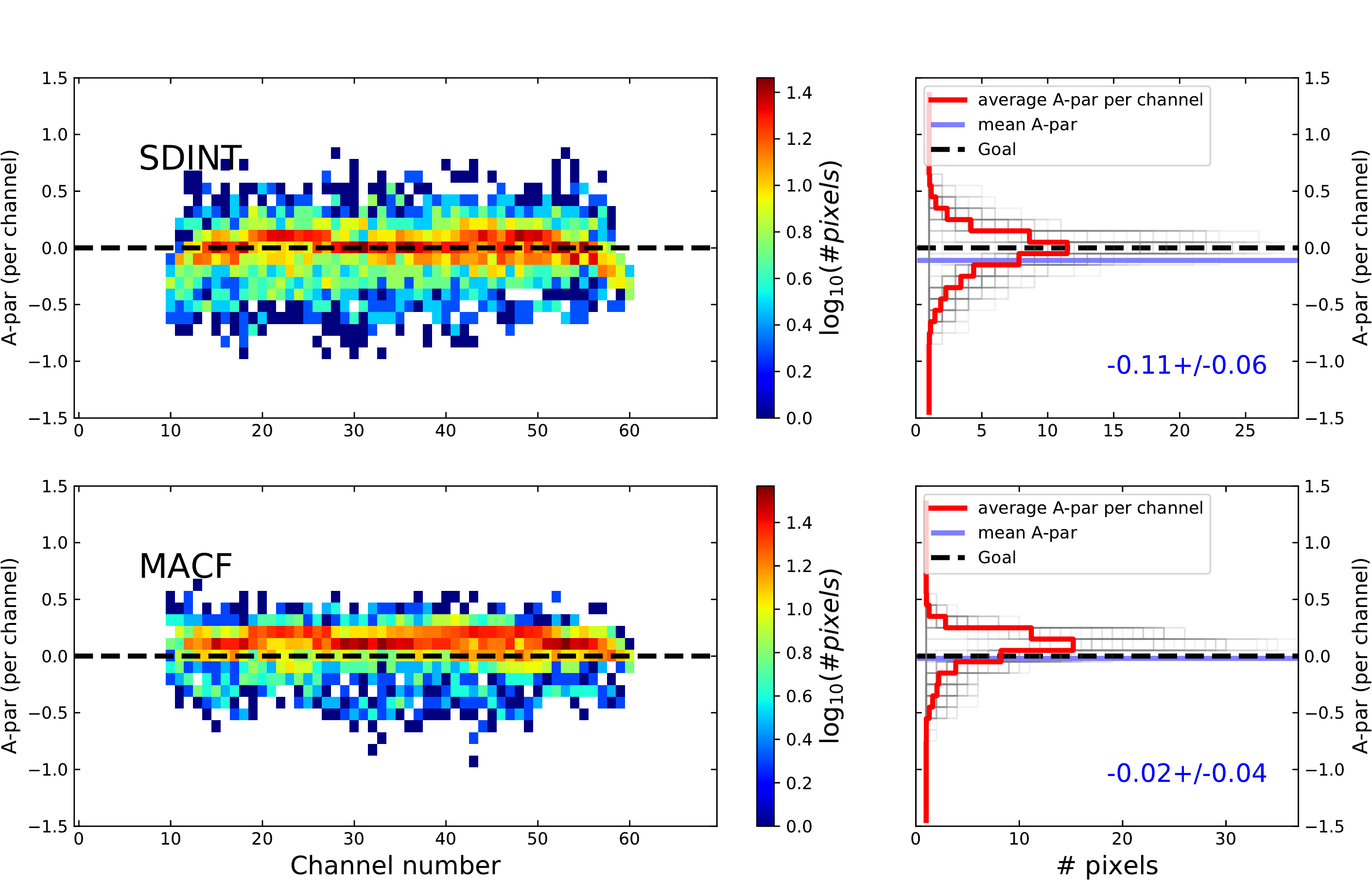}
\caption{A-par spectrograms for SDINT (upper panel) and MACF (lower panel) methods for the case of M100. See also Fig.~\ref{fig:m100_apar}.    \label{fig:m100_spectrogram_extended}}
\end{figure}

In \S~\ref{sec:comp-w-sd} we explained that for actual observational data, the SD image needs to serve as the reference image. Fig. \ref{fig:m100_apar} already showed the A-par spectrograms for the int-CLEAN image cube and the Feather combined image cube of our M100 data using the M100 SD image cube as reference. In Fig. \ref{fig:m100_spectrogram_extended} we show the corresponding plots for the remaining combination methods SDINT and MACF. Similar to the case of our continuum maps (\S~\ref{sec:short_spacing}), all data combination methods produce significant improvements in the image quality of the spectral cube. 

In the case of our M100 dataset all methods produce satisfactory results with deviations from the expected flux per channel reduced to $\lesssim$10\% in all cases.
The best results are produced by Feather and MACF methods both showing narrow A-par distributions ($\sigma($~A-par$)=0.04$) with an average flux recovery of $\gtrsim$~95\% ($|$A-par$| <$~0.05) while SDINT shows slightly larger differences with respect to the SD flux ($|$A-par$|=-0.11$ and $\sigma($~A-par$)=0.06$). Comparisons within these 10\% variations are  however difficult to interpret given the low number statistics available ($\sim$~20~voxels per channel within the assessment mask) once these analyses are carried out at the SD resolution.

Compared with the results obtained on our Skymodel, where SDINT and MACF produced the best reductions (\S~\ref{sec:method_comparison}), our analysis suggests that the performance of each combination method may slightly vary depending on spatial structure of the astronomical target, which may vary between velocity channels; \reply{there may be additional dependence on} the depth and type of observation. While our analysis does not allow to establish the clear superiority of any of the data combination methods in all situations, our assessment metrics appear to be a useful diagnostic toolkit to systematically investigate and compare their performance, statistically quantify their ability to recover the true sky emission at different scales, and identify potential combination issues in both continuum and spectral datasets.
For simplicity, our data combinations assumed standard values for reduction parameters such as \texttt{briggs} (CLEAN), \texttt{sdfactor} (Feather and MACF) or \texttt{sdgain} (SDINT), among others (see \S~\ref{sec:imaging} and \ref{sec:methods}).
The assessments presented here can also be extended to quantify the effect and sensitivity on the flux and scales recovered by other parameter choices and techniques. Users of these methods are therefore recommended to evaluate their results using these diagnostics.

\section{Summary}

In this paper we underline with synthetic and real-world examples the importance of the combination of single-dish (SD) data with interferometric data in radio astronomy when the objects observed emit on scales that are larger than a few times the angular resolution of the interferometric observation. We then briefly describe the prominent data combination methods available today (Feather, TP2VIS, MACF, SDINT, FSSC) and test them on two datasets, one simulated and one observational, with a new set of tools producing several novel diagnostic plots which permit one to quantify and compare the performance of the combination methods. These tools are mostly based on the accuracy parameter, ``A-par," which is essentially the relative difference between the output image of the method and the original input image of the simulation (in the case of simulated test data) or the original SD image (in the case of a real observation). We also discuss the commonly used Fidelity parameter, and we find A-par to have higher diagnostic value.
An additional valuable diagnostic (closely related to A-par) is the fraction of recovered flux as a function of reference flux either per reference flux bin or cumulative. 

For a given observation, the A-par can be inspected in four different ways: as a spatial map, as a histogram of the A-par value for all pixels, as a power spectrum of the A-par map (as a function of angular scale), and for spectrally resolved data (image cubes) as a spectrogram which shows the per-channel A-par histograms for each channel next to each other. 
For individual images (a single spectral channel) the A-par maps and histograms can clearly reveal the need for data combination.
For spectrally resolved data, we demonstrate that not applying data combination can result in significantly deformed spectral line shapes and intensity ratios and this is clearly visible in the A-par spectrograms.  

As a result of our analysis, we recommend obtaining additional SD observations for all interferometric observations of significantly resolved objects \reply{amidst extended emission,} and to produce diagnostic A-par plots in order to verify (a) whether combination with SD data is required and (b) whether the combination was performed optimally.

Our results on the performance of the tested data combination methods show \reply{quantitatively} that the application of any method is better than using the interferometric data on its own. Interferometric-alone reductions can miss up to 90\% (A-par=-0.9) of the flux depending on the scale and source morphology. These flux losses are present at all scales, affecting both the diffuse emission and compact sources down to scales similar to the beamsize. In contrast, the use of data combination techniques such as Feather, SDINT, or MACF allows the recovery of most of the true sky emission within errors of $\lesssim$~10\% ($|$A-par$|<$~0.1). Our tests show that the enhanced performance of some of these methods under certain circumstances can reduce the mean value ($|$A-par$|<$~0.05) and variance ($\sigma($A-par$)<$~0.05) of these differences to $<$~5\% with respect to the actual target flux.


All combination methods have a parameter which permits tuning the contribution of the SD data, the choice of which depends on the relative sensitivity achieved in the interferometric and SD observations. While this parameter should be chosen on first principles, \reply{one can also tune the relative weighting of the input data depending on their sensitivities and the scientific goal of the experiment.} Moreover, the goodness of these different combinations can be further quantified with our assessment diagnostics.

Recovering the true sky emission at all scales has fundamental implications for the physical interpretation of scientific images. For instance, the continuum flux is directly connected to measurements of the (gas and dust) column densities, probability distribution functions, spectral indices, and masses. Similarly, the line emission profiles provide information on the gas excitation conditions, molecular abundances, or gas chemistry, among others. 
The addition of the zero-spacing information is revealed as fundamental to produce accurate, high-dynamic range images in radio-astronomical, interferometric observations of spatially resolved targets with complex emission substructure.
Our results demonstrate the importance of implementing advanced data combination techniques in current (e.g., NOEMA, SMA, and ALMA) and future (e.g., ngVLA and SKA) interferometric studies.
The reader is encouraged to try our demonstration software package available from \packageurl.

\acknowledgments

The authors gratefully acknowledge the thorough review and constructive comments from the very knowledgeable anonymous referee. The authors acknowledge the Lorentz Center, which enabled the launch of this initiative through funding, organizing and hosting the workshop ``Improving Image Fidelity on Astronomical Data: Radio Interferometer and Single-Dish Data Combination'' held on 12 - 16 August 2019. The European Southern Observatory provided additional support. We particularly thank Hauyu Liu, Yanett Contreras, and Alvaro Sanchez-Monge for co-organizing this 2019 Lorentz Center workshop.  Including, and in addition to, participants of the Lorentz Center workshop, we acknowledge in particular: Amanda Kepley, Urvashi Rau, Sandra Burkutean, Adam Ginsburg, and Jens Kauffmann, who contributed to development and discussion of techniques used in this manuscript.  This paper makes use of the following ALMA data: ADS/JAO.ALMA\#2011.0.00004.SV. ALMA is a partnership of ESO (representing its member states), NSF (USA) and NINS (Japan), together with NRC (Canada), MOST and ASIAA (Taiwan), and KASI (Republic of Korea), in cooperation with the Republic of Chile. The Joint ALMA Observatory is operated by ESO, AUI/NRAO and NAOJ. The National Radio Astronomy Observatory is a facility of the National Science Foundation operated under cooperative agreement by Associated Universities, Inc.
AH and SS acknowledge the support and funding from the European
Research Council (ERC) under the European Union’s Horizon 2020
research and innovation programme (Grant agreement Nos. 851435).
AH acknowledges assistance from Allegro, the European ALMA Regional Center node in the Netherlands.
LMF acknowledges funding by the German Federal Ministry of Education and Research under the funding code 05A20PD1.
YM acknowledges support from grant JSPS KAKENHI (Grant Number JP20K04034). KMH acknowledges financial support from the grant SEV-2017-0709 funded by MCIN/AEI/ 10.13039/501100011033, from the coordination of the participation in SKA-SPAIN, funded by the Ministry of Science and Innovation (MCIN), and financial support from the grants RTI2018-096228-B-C31 and PID2021-123930OB-C21 funded by MCIN/AEI/ 10.13039/501100011033, by ``ERDF A way of making Europe'' and by the ``European Union''.
This research made use of Astropy (http://www.astropy.org), a community-developed core Python package for Astronomy \citep{astropy2013,astropy2018}.
\vspace{5mm}

\facilities{ALMA}

\software{CASA \citep{casa2022},
  astropy \citep{astropy2013,astropy2018},
  numpy \citep{harris2020array},
  scipy \citep{2020SciPy-NMeth},
  matplotlib \citep{Hunter:2007},
  analysisUtils (\url{https://casaguides.nrao.edu/index.php/Analysis_Utilities})
}


\bibliography{biblio}{}
\bibliographystyle{aasjournal}

\end{document}